\documentclass[11pt,reqno]{amsart}%{article}
\usepackage{amssymb}
\newtheorem{theorem}{Theorem}[section]
\newtheorem{proposition}{Proposition}[section]

\newtheorem{lemma}{Lemma}[section]
\newtheorem{definition}{Definition}[section]

\newfont{\bb}{msbm10 at 12pt}

\newcommand{\mysection}[1]{\section{#1}\setcounter{equation}{0}}

\def\pf{{\textit {Proof :} }}

\def\R{\hbox{\bb R}}
\newcommand{\bal}{\begin{align}}     \newcommand{\eal}{\end{align}}
\newcommand{\ba}{\begin{array}}      \newcommand{\ea}{\end{array}}
\newcommand{\bc}{\begin{center}}     \newcommand{\ec}{\end{center}}
\newcommand{\be}{\begin{enumerate}}  \newcommand{\ee}{\end{enumerate}}
\newcommand{\beq}{\begin{eqnarray}}  \newcommand{\eeq}{\end{eqnarray}}
\newcommand{\beQ}{\begin{eqnarray*}} \newcommand{\eeQ}{\end{eqnarray*}}
\newcommand{\bi}{\begin{itemize}}    \newcommand{\ei}{\end{itemize}}
\newcommand{\bt}{\begin{tabular}}    \newcommand{\et}{\end{tabular}}
\newcommand{\bdm}{\begin{displaymath}} \newcommand{\edm}{\end{displaymath}}

\let\pa=\partial

\def\qed{\hfill{Q.E.D.}\smallskip}
\newcommand{\ls}{\setlength{\baselineskip}{12pt}
                 \setlength{\parskip}{3mm}}

\usepackage{upgreek}
\usepackage{tensor}
\usepackage{mathrsfs}

\begin{document}

% Jialue Li add command, make the formula split into different pages
\allowdisplaybreaks

\title[Bondi-Sachs Formalism]
{The Bondi-Sachs Formalism for the Einstein Scalar Field Equations with the Zero Cosmological Constant}

\author[J Li]{Jialue Li$^{1,3}$}
\author[X Zhang]{Xiao Zhang$^{1,2,4}$}

\address[]{$^{1}$Academy of Mathematics and Systems Science, Chinese Academy of Sciences, Beijing 100190, People's Republic of China}
\address[]{$^{2}$School of Mathematical Sciences, University of Chinese Academy of Sciences, Beijing 100049, People's Republic of China}
\address[]{$^{3}$School of Mathematical Sciences, Peking University, Beijing 100871, People's Republic of China}
\address[]{$^{4}$Guangxi Center for Mathematical Research, Guangxi University, Nanning 530004, Guangxi, People's Republic of China}

\email{lijialue@amss.ac.cn$^{1}$, lijialue@pku.edu.cn$^{3}$}
\email{xzhang@amss.ac.cn$^{1,2}$, xzhang@gxu.edu.cn$^{4}$}

\begin{abstract}
Inspired by interaction of gravitational waves and dark matters, we study the Bondi-Sachs formalism for Einstein massless scalar field with zero cosmological constant. We provide asymptotic expansions for the Bondi-Sachs metrics as well as the scalar fields and prove the peeling property. We also prove the positivity of the Bondi energy-momentum under condition $c=d=0$ at some retarded time $u_0$. This condition ensures that asymptotically null hypersurfaces near $u=u_0$ are asymptotically null initial data sets of order 2 and the positive energy theorem for null infinity can be applied.
\end{abstract}

%\thanks{}

\date{}

\maketitle \pagenumbering{arabic}
\pagenumbering{arabic}
%%%%%%%%%%%%%%%%%%%%%%%%%%%%%%%%%%%%%%%%%%%%%%%%%%%%%%%%%%%%%%%%%%%%%%%%%%%%%

\mysection{Introduction}
\ls

In general relativity, a spacetime is a four-dimensional Lorentzian manifold whose metric ${\bf g}$ satisfies the Einstein field equations
\begin{equation}\label{eq:EinsteinTotal}
    R_{\mu\nu}-\frac{R}{2}{\bf g}_{\mu\nu}+\Lambda {\bf g}_{\mu\nu}=T_{\mu\nu},
\end{equation}
where $R_{\mu\nu}$ is the Ricci curvature, $R$ is the scalar curvature, $\Lambda$ is the cosmological constant and $T_{\mu\nu}$ is the energy-momentum tensor of matter. In particular, it is a vacuum when $T_{\mu\nu}$ vanishes. When it is wave-like
and radiates energy, the metric ${\bf g}$ is referred to as gravitational waves, see, e.g., \cite{Z6}. The existence of gravitational waves was
predicted by Einstein theoretically and was first detected by LIGO and Virgo in September, 2015, which is a black hole event, named
GW150914 \cite{Aetc}. Measured by solar mass, the initial black hole masses are 36 units and 29 units, and they collide and form the final
black hole whose mass is 62 units, with 3 units of mass radiated away in gravitational waves.

Gravitational waves are assumed to be vacuum spacetimes. When the cosmological constant is zero, they were firstly studied by Bondi, van der Burg, Metzner for axially symmetric isolated spacetimes and by Sachs for general asymptotically flat spacetimes in 1962 \cite{BBM, S}. They introduced the so-called Bondi-Sachs metrics 
\beq
\begin{aligned}\label{eq:BondiSachsMetric}
{\bf g}=&-\bigg( \frac{\mathrm{e}^{2\beta}V}{r} -r^2\Big(\mathrm{e}^{2\gamma}U^2\cosh(2\delta)+2 UW\sinh(2\delta) \\
              &+\mathrm{e}^{-2\gamma}W^2\cosh(2\delta)\Big)\bigg)\mathrm{d}u^2-2\mathrm{e}^{2\beta}\mathrm{d}u\mathrm{d}r \\
              & -2r^2\bigg(\mathrm{e}^{2\gamma}U\cosh(2\delta)+W\sinh(2\delta)\bigg)\mathrm{d}u\mathrm{d}\theta \\
	          & -2r^2\bigg(\mathrm{e}^{-2\gamma}W\cosh(2\delta)+U\sinh(2\delta)\bigg)\sin\theta \mathrm{d}u\mathrm{d}\phi \\
	          & +r^2\bigg(\mathrm{e}^{2\gamma}\cosh(2\delta)\mathrm{d}\theta^2+2\sinh(2\delta)\sin\theta\mathrm{d}\theta \mathrm{d}\phi \\
              & +\mathrm{e}^{-2\gamma}\cosh(2\delta)\sin^2\theta\mathrm{d}\phi^2\bigg)
\end{aligned}
\eeq
in coordinates $\{u,\,r,\,\theta,\,\phi\}$ ($u$ is retarded time)
\begin{equation*}
    -\infty<u<\infty,\quad r\geqslant 0,\quad 0\leqslant \theta\leqslant \pi,\quad
    0\leqslant \phi\leqslant 2\pi,
\end{equation*}
where $\beta$, $\gamma$, $\delta$, $U$, $V$ and $W$ are smooth functions of $u$, $r$, $\theta$ and $\phi$, which are defined on $S^2$ for each $u$, i.e., they and their derivatives take the same value at $\phi=0, \,2\pi$. It assumes that \eqref{eq:BondiSachsMetric} is asymptotically flat, i.e., for sufficiently large $r$, it takes the form:
\begin{align}
{\bf g}=-\mathrm{d}u^2-2\mathrm{d}u\mathrm{d}r
	          +r^2\big(\mathrm{d}\theta^2+\sin^2\theta\mathrm{d}\phi^2\big)+O\left(\frac{1}{r}\right).\label{eq:BSmetric1}
\end{align}
By assuming the {\it outgoing radiation condition} 
\begin{align*}
\gamma = \frac{c(u, \theta, \phi)}{r} +O\left(\frac{1}{r^3}\right), \quad \delta =\frac{d(u, \theta, \phi)}{r} +O\left(\frac{1}{r^3}\right),
\end{align*}
the vacuum Einstein field equations imply
\begin{align*}
V=r-2M(u, \theta, \phi)+O\left(\frac{1}{r}\right).
\end{align*}
Physically, $c_u$, $d_u$ are referred to as {\it news functions}, and $M$ is referred to as {\it mass aspect} of the vacuum Einstein field equations for \eqref{eq:BondiSachsMetric}. Denote
\begin{equation*}
%\label{eq:nnuDef}
    n^0=1,\quad n^1=\sin\theta\cos\phi,\quad n^2=\sin\theta\sin\phi,\quad n^3=\cos\theta.
\end{equation*}
Bondi defined the Bondi energy-momentum of each null hypersurface as
\begin{align*}
    m_\nu(u)=\frac{1}{4\pi}\int_{S^2}M(u,\theta,\phi)n^\nu\mathrm{d}S, \quad \nu=0, 1, 2, 3,
\end{align*}
and derived its non-increasing property 
\begin{align*}
    \frac{\mathrm{d}}{\mathrm{d}u} m_0(u) \leqslant 0
\end{align*}
under the following condition \cite{HYZ}
\begin{align}
    \int_0 ^\pi c(u, 0, \phi)\mathrm{d}\phi=0, \quad \int_0 ^\pi c(u, \pi, \phi)\mathrm{d}\phi=0 \label{hyz-u}
\end{align}
for any $u$. Soon later, Penrose introduced simple spacetimes to study gravitational waves in terms of conformal compactification and showed the peeling property of the Weyl curvatures, see, e.g., \cite{NP1, NP2, NU}. The Bondi energy is referred to as the total energy after loss carried away by gravitational waves. It is a fundamental problem whether isolated gravitational systems can radiate away more energy than they initially have, i.e., whether Bondi's energy is nonnegative. The proofs of this positivity for vacuum spacetimes were claimed by using both Schoen-Yau and Witten's arguments of proving the positive energy theorem for asymptotically flat initial data sets, c.f. \cite{SY, CJL} and references therein, but completed under some extra assumptions in \cite{HYZ} via the Dirac operator method, see also \cite{Z4, Z5}. The key point is to apply the positive energy theorem for null infinity, which was proved in \cite{Z1} by using the Dirac operator method. This theorem holds for asymptotically null initial data sets where both metrics and the second fundamental forms of hypersurfaces are asymptotically hyperbolic with order greater than $\frac{3}{2}$ in asymptotically flat spacetimes satisfying the dominant energy condition. However, asymptotically null initial data sets in vacuum Bondi-Sachs metrics are always of order 1 \cite{HYZ} (see, also \S 4). Therefore, we need extra condition $c=d=0$ at some retarded time $u_0$ to ensure asymptotically null hypersurfaces near $u=u_0$ are asymptotically null initial data sets of order 2 and the positive energy theorem for null infinity can be applied \cite{HYZ, Z4, Z5}. 

Denote 
\begin{align}
    l = c_\theta+2c\cot\theta+d_\phi\csc\theta, \quad
    \hat{l} =d_\theta+2d\cot\theta-c_\phi\csc\theta, \label{l-barl}
\end{align}
and
\begin{equation}\label{eq:NewMandOldMrelation}
    \mathcal{M}(u, \theta, \phi)=M(u, \theta, \phi)-\frac{1}{2}\left(l_\theta+l\cot\theta+\hat{l}_\phi\csc\theta\right).
\end{equation}
($\mathcal{M}$ is referred to as the {\it modified mass aspect} of the vacuum Einstein field equations for Bondi-Sachs metric \eqref{eq:BondiSachsMetric} \cite{HYZ}.) Let $0<\varepsilon<1$. For Schoen-Yau's argument, it was proposed to solve the Jang equation for some $u_0$ with asymptotics \cite{SY}
\begin{align*}
f(r, \theta, \phi) =\sqrt{1+r^2} +2 \mathcal{M}(u_0, \theta, \phi) \ln r +O\left(\frac{1}{r^{1-\varepsilon}}\right).
\end{align*}
But this is restricted and requires $\mathcal{M}$ to be constant \cite{HYZ}. The suitable asymptotics 
\begin{align*}
f(r, \theta, \phi) =\sqrt{1+r^2} +2 \mathcal{M}(u_0, \theta, \phi) \ln r +\tilde{f}(\theta, \phi)+O\left(\frac{1}{r^{1-\varepsilon}}\right)
\end{align*}
was provided by Sakovich \cite{Sak} and the Jang equation was solved over asymptotically null hyperbolic hypersurfaces with Wang's asymptotics, which both metrics and the second fundamental forms are of the form
\begin{align*}
    \frac{\mathrm{d}r^2}{1+r^2}+r^2 \left(\mathrm{d}\theta^2+\sin^2\theta\mathrm{d}\phi^2+O\left(\frac{1}{r^3}\right)\right).
\end{align*}
Wang's asymptotics is much stronger than the usual asymptotical hyperbolicity \cite{W, CH, Z1}. But, if it satisfies the dominant energy condition, then an asymptotically null initial data set of order greater than $\frac{3}{2}$ can be deformed to another asymptotically null initial data set with Wang's asymptotics and slight different mass functionals, which satisfies the strict inequality in the dominant energy condition \cite{DS}. In this case, for null infinity, positivity of the energy can be proved for asymptotically null initial data sets, but result for the vanishing energy holds only for those with Wang's asymptotics. If $c=d=c_u=d_u=0$ at some retarded time $u_0$, then asymptotically null hypersurfaces near $u=u_0$ satisfy conditions in \cite{Sak} and the positivity of the Bondi energy holds. 

Physically, positivity of the Bondi energy should require conservation of gravitational system's ADM total energy-momentum. However, asymptotics of Bondi-Sachs metrics is weak and not sufficient for the ADM total energy-momentum to be conserved. Therefore, the Bondi energy might be negative in general and the conjecture for positivity of the Bondi energy-momentum might not be true.

For other issues related to the Bondi-Sachs spacetimes, we refer to Wang \cite{WangMT} and the reference therein for study of the angular momentum, and to e.g., \cite{GLSWZ, XZ} for study of the case of nonzero cosmological constant. We point out that, when the cosmological constant is nonzero, some cosmological constraints occur for the boundary conditions. This causes the situation on asymptotics is completely different, and the good definition for the Bondi energy is still lacking.

In this paper, we study the Bondi-Sachs formulism for the Einstein scalar field equations when the cosmological constant is zero. Physically, scalar fields can describe dark matter, see, e.g., \cite{AGM}. Therefore, the theory can be referred to as interaction of gravitational waves and dark matters. Denote by $\nabla$ the Levi-Civita connection of ${\bf g}$. For massless scalar field $\Psi$, the energy-momentum tensor is given by
\begin{equation}\label{eq:scalarTDef}
    T_{\mu\nu}=\nabla_\mu\Psi\nabla_\nu\Psi-\frac{1}{2}{\bf g}_{\mu\nu}\nabla^\sigma\Psi\nabla_\sigma\Psi.
\end{equation}
The Einstein scalar field equations are
\begin{equation}
    R_{\mu\nu}=\nabla_\mu\Psi\nabla_\nu\Psi. \label{esf}
\end{equation}
Denote $\Box=\nabla ^\mu \nabla_\mu$ as the wave operator. Twice contracted Bianchi identity implies that
\begin{equation}
\Box\Psi =0. \label{wave equation}
\end{equation}
We assume also that the outgoing radiation condition holds. If the scalar field $\Psi$ satisfies
\begin{align*}
\Psi = O\left(\frac{1}{r}\right),
\end{align*}
then the Einstein scalar field equations imply
\begin{align*}
   V =r-2\mathcal{M}(u, \theta, \phi)-\big(l_\theta+l\cot\theta+\hat{l}_\phi\csc\theta\big)+O\left(\frac{1}{r}\right).
\end{align*}
We follow from \S 6 in \cite{HYZ} to modify Bondi's definition by using $\mathcal{M}$ to define the Bondi energy-momentum of each null hypersurface as
\begin{equation}\label{eq:newMBondiEnergyMomentumDef}
    m_\nu(u)=\frac{1}{4\pi}\int_{S^2}\mathcal{M}(u,\theta,\phi)n^\nu\mathrm{d}S, \quad \nu=0, 1, 2, 3.
\end{equation}
For abusing titles, we call $m_\nu$, $\mathcal{M}$ also the Bondi energy-momentum and {\it mass aspect}. The advantage is that the new Bondi energy is non-increasing with respect to $u$ without assuming the unnatural condition \eqref{hyz-u}. The outgoing radiation condition and decaying of $\Psi$ ensure that, if $c=d=0$ at some retarded time $u_0$, asymptotically null hypersurfaces near $u=u_0$ are asymptotically null initial data sets of order 2 (c.f. \S 5). Thus, we can prove the positivity of the Bondi energy-momentum for Einstein scalar field equations by using the positive energy theorem for null infinity (Theorem \ref{main thm}).

This paper is organized as follows.
In Sect. 2, we study the structure of the Einstein scalar field equations for Bondi-Sachs metrics and separate them into seven equations.
In Sect. 3, we provide asymptotic expansions for the Bondi-Sachs metrics as well as the scalar fields.
In Sect. 4, we prove the peeling property for the Einstein scalar field equations.
In Sect. 5, we derive the first and the second fundamental forms of asymptotically null spacelike hypersurfaces in Bondi-Sachs spacetimes and show that they are asymptotically null initial data sets of order 1 in general.
In Sect. 6, we prove the positivity of the Bondi energy-momentum for the Einstein scalar fields under condition $c=d=0$ at some retarded time $u_0$. This condition ensures that asymptotically null hypersurfaces near $u=u_0$ are asymptotically null initial data sets of order 2 and the positive energy theorem for null infinity can be applied.
In Appendix A, we provide explicit formulas relating to the Einstein scalar field equations.
In Appendix B, we provide explicit formulas for certain higher order coefficients in asymptotic expansions in Sect. 3.

\mysection{Einstein scalar field equations for Bondi-Sachs metrics}
\ls

In this section, we study the structure of the Einstein scalar field equations for Bondi-Sachs metric \eqref{eq:BondiSachsMetric}.

\begin{definition}
A four-dimensional Lorentzian manifolds ($\mathscr{L}$, ${\bf g}$) is called an asymptotically flat Bondi-Sachs spacetime if there exists a four-dimensional Lorentzian manifold $\mathscr{L}_c$ foliated by compact three-dimensional spacelike hypersurfaces and a diffeomorphism
\begin{align*}
\mathscr{L} \backslash \mathscr{L}_c   \longmapsto \R ^{3,1} \backslash (B^3 _{R_0}\times \R)
\end{align*}
for some $R_0 >0$, where $B^3 _{R_0}$ is a central ball in $R^3$ with radius $R_0$, and the metric ${\bf g}$ takes the form \eqref{eq:BondiSachsMetric} with asymptotic expansion \eqref{eq:BSmetric1} on $\mathscr{L} \backslash \mathscr{L}_c$.
\end{definition}

\begin{proposition}
Let $\Psi$ be any smooth scalar field over a Bondi-Sachs spacetime. Denote
\begin{align}
\Omega_{\mu\nu} =R_{\mu\nu}-\nabla_\mu\Psi\nabla_\nu\Psi, \quad \Omega={\bf g}^{\mu\nu}\Omega_{\mu\nu}. \label{omega}
\end{align}
It holds that
\begin{align}
	\nabla^\mu\bigg(\Omega_{\mu\nu}-\frac{\Omega}{2}{\bf g}_{\mu\nu}\bigg)
	=-\Box\Psi \nabla_\nu \Psi.  \label{omega1}
\end{align}
\end{proposition}
\pf By the twice contracted Bianchi identity,
\begin{align*}
{\rm LHS}
=&\nabla^\mu\bigg(R_{\mu\nu}-\frac{R}{2}{\bf g}_{\mu\nu}-\nabla_\mu\Psi\nabla_\nu\Psi+\frac{\nabla^\sigma\Psi\nabla_\sigma\Psi}{2} {\bf g}_{\mu\nu}\bigg) \\
=&-\Box\Psi \nabla_\nu \Psi -\nabla_\mu\Psi\nabla^\mu\nabla_\nu\Psi +\nabla_\nu \nabla^\sigma\Psi \nabla_\sigma\Psi \\
    =&-\Box\Psi \nabla_\nu \Psi- \nabla_\mu\Psi\big(\nabla^\mu\nabla_\nu\Psi -\nabla_\nu \nabla^\mu\Psi \big)\\
=&-\Box\Psi \nabla_\nu \Psi.
\end{align*}
Therefore the proposition holds. \qed

The Einstein scalar field equations \eqref{esf} give that
\begin{equation}\label{eq:EinsteinTotalOmega}
    \Omega_{\mu\nu}=0
\end{equation}
for $\mu,\,\nu=0,\,1,\,2,\,3$, where we denote coordinates with indices
\begin{align*}
x^0=u,\quad x^1=r, \quad x^2=\theta, \quad x^3=\phi.
\end{align*}
Same as \cite{BBM, S, vdB}, these ten equations are separated into three groups
\begin{enumerate}
   \item six main equations
\begin{equation}\label{eq:main6eq}
    \Omega_{11}=\Omega_{12}=\Omega_{13}=\Omega_{22}=\Omega_{23}=\Omega_{33}=0,
\end{equation}
   \item one trivial equation
\begin{equation}\label{eq:trivial1eq}
    \Omega_{01}=0,
\end{equation}
   \item three supplementary equations
\begin{equation}\label{eq:supplementary3eq}
    \Omega_{00}=\Omega_{02}=\Omega_{03}=0.
\end{equation}
\end{enumerate}
For Bondi-Sachs metric \eqref{eq:BondiSachsMetric}, 
\begin{align}
{\bf g}_{11}={\bf g}_{12}={\bf g}_{13}=0 \Longrightarrow {\bf g}^{00}={\bf g}^{02}={\bf g}^{03}=0. \label{g-zero},
\end{align}
and the Christoffel symbols of the Bondi-Sachs metrics give that \cite{BBM, S, vdB}
\begin{equation}\label{eq:Christoffel0g}
    {\bf g}^{\alpha\epsilon} \Gamma^0_{\alpha\epsilon}=\frac{2\mathrm{e}^{-2\beta}}{r}
\end{equation}
and \eqref{esf}, \eqref{omega1} give that
\begin{equation}\label{eq:gChristoffelBianchi}
    {\bf g}^{\alpha\epsilon}
	\left( \frac{\partial \Omega_{\mu\alpha}}{\partial x^\epsilon}
	-\frac{1}{2}\frac{\partial \Omega_{\alpha\epsilon}}{\partial x^\mu}
	-\Gamma^\delta _{\alpha\epsilon} \Omega_{\mu\delta}\right)=0
\end{equation}
for $\mu=0,\,1,\,2,\,3$. 

In the following, we provide details that the six main equations \eqref{eq:main6eq} imply that the trivial equation \eqref{eq:trivial1eq} holds. Moreover, they imply that the three supplementary equations hold everywhere if they hold for at some $r_0 >0$ \cite{BBM, S, vdB}. Indeed, if we assume that \eqref{eq:main6eq} holds, then
\begin{enumerate}
   \item $\mu=1$: using \eqref{eq:main6eq} and \eqref{g-zero}, \eqref{eq:gChristoffelBianchi} gives
\begin{equation}\label{omega01}
{\bf g}^{01} \frac{\partial \Omega _{10}}{\partial r}-\frac{{\bf g}^{01}}{2}\frac{\partial \Omega _{01}}{\partial r}-\frac{{\bf g}^{10}}{2}\frac{\partial \Omega _{10}}{\partial r}-{\bf g}^{\alpha\epsilon}\Gamma^0_{\alpha\epsilon}\Omega_{10}=-{\bf g}^{\alpha\epsilon}\Gamma^0_{\alpha\epsilon}\Omega_{01}=0.\\
\end{equation}
%\begin{equation}\label{omega01}
%{\bf g}^{01} \frac{\partial \Omega _{01}}{\partial r}-\frac{{\bf g}^{01}}{2}\frac{\partial \Omega _{01}}{\partial r}-\frac{{\bf g}^{10}}{2}\frac{\partial \Omega _{10}}{\partial r}-{\bf g}^{\alpha\epsilon}\Gamma^0_{\alpha\epsilon}\Omega_{01}=-{\bf g}^{\alpha\epsilon}\Gamma^0_{\alpha\epsilon}\Omega_{01}=0.\\
%\end{equation}
    \item $\mu=A$: using \eqref{eq:trivial1eq} and \eqref{eq:Christoffel0g}, \eqref{eq:gChristoffelBianchi} gives
\begin{equation}\label{eq:Omega0A}
    \frac{\mathrm{e}^{-2\beta}}{r^2}\frac{\partial}{\partial r}\left(r^2\Omega_{0A}\right) =0,\quad A=2,3.\\
\end{equation}
    \item $\mu=0$: using \eqref{eq:trivial1eq}, \eqref{eq:Christoffel0g} and
        $\Omega_{02}=\Omega_{03}=0$, \eqref{eq:gChristoffelBianchi} gives
\begin{equation}\label{eq:Omega00}
    \frac{\mathrm{e}^{-2\beta}}{r^2}\frac{\partial}{\partial r}\left(r^2\Omega_{00}\right) =0.
\end{equation}
\end{enumerate}
Therefore, \eqref{eq:trivial1eq} holds by \eqref{eq:Christoffel0g}, and \eqref{omega01}, \eqref{eq:supplementary3eq} hold everywhere if they hold at some $r_0 >0$ by \eqref{eq:Omega0A}, \eqref{eq:Omega00}. So it only needs to study the six main equations, which separate into two groups:
\begin{enumerate}
    \item four hypersurface equations
        \[ \Omega_{11}=\Omega_{12}=\Omega_{13}=0, \]
	\[ \left(\mathrm{e}^{-2\gamma}\Omega_{22}
	+\mathrm{e}^{2\gamma}\csc^2\theta\Omega_{33}\right)
	\cosh(2\delta)-2\csc\theta \Omega_{23}\sinh(2\delta)=0;  \]
    \item two standard equations
	\[ \mathrm{e}^{-2\gamma}\Omega_{22}-\mathrm{e}^{2\gamma}\csc^2\theta\Omega_{33}=0,  \]
	\[ \left(\mathrm{e}^{-2\gamma}\Omega_{22}
	+\mathrm{e}^{2\gamma}\csc^2\theta\Omega_{33}\right)\sinh(2\delta)
	-2\csc\theta \Omega_{23}\cosh(2\delta)=0.  \]
\end{enumerate}
We conclude that the Einstein scalar field equations \eqref{esf} are equivalent to the following seven equations
\begin{align}
    \beta_r &=\mathscr{R}_1,  \label{eq:EQ1} \\
    \left(r^4\mathrm{e}^{-2\beta}\left(\mathrm{e}^{2\gamma}U_r\cosh(2\delta)+W_r\sinh(2\delta) \right)\right)_r
    &=\mathscr{R}_2, \label{eq:EQ2} \\
    \left(r^4\mathrm{e}^{-2\beta}
    \left(U_r\sinh(2\delta)+\mathrm{e}^{-2\gamma}W_r\cosh(2\delta)\right)\right)_r &=\mathscr{R}_3,  \label{eq:EQ3} \\
     V_r &=\mathscr{R}_4, \label{eq:EQ4} \\
       (r\gamma)_{ur}\cosh(2\delta)
    +2r(\gamma_u\delta_r+\delta_u\gamma_r)\sinh(2\delta)
    & =\mathscr{R}_5,  \label{eq:EQ5} \\
      (r\delta)_{ur}-2r\gamma_u\gamma_r\sinh(2\delta)\cosh(2\delta)
    &=\mathscr{R}_6,  \label{eq:EQ6} \\
     (r\Psi_u)_{r} &=\mathscr{R}_7, \label{eq:EQ7}
\end{align}
where $\mathscr{R}_1,\ldots,\mathscr{R}_7$ are given in Appendix A. The most important feature is that $\mathscr{R}_1,\ldots,\mathscr{R}_7$ do not contain any derivatives with respect to $x^0=u$.

{\em Null-timelike boundary problems for \eqref{eq:EQ1}-\eqref{eq:EQ7}: given initial data
\begin{align*}
\gamma | _{u=u_0}, \quad \delta | _{u=u_0}, \quad \Psi | _{u=u_0}
\end{align*}
on a null hypersurface $\{u=u_0\}$, $-\infty <u_0<\infty$, boundary values
\begin{align*}
(r^4 U_r) | _{r=r_0},\quad (r^4 W_r)| _{r=r_0}, \quad V| _{r=r_0}
\end{align*}
on timelike hypersurface $\{r=r_0\}$, $0\leqslant r_0 <\infty$, and
\begin{align*}
\beta | _{r=r_1},\quad U| _{r=r_1} ,\quad W| _{r=r_1},\quad (r\gamma _u) | _{r=r_1}, \quad (r \delta_u) | _{r=r_1}, \quad (r\Psi_u) | _{r=r_1} 
\end{align*}
on timelike hypersurface $\{r=r_1\}$, $0 \leqslant r_1 \leqslant \infty$, such that \eqref{eq:supplementary3eq} holds on $\{r=r_0\}$ and $\{r=r_1\}$, which could be viewed as constraint equations, can one prove the existence and uniqueness of \eqref{eq:EQ1}-\eqref{eq:EQ7}?}

We refer to \cite{HZ} on energy estimates and existence for null-timelike boundary problems of linear wave equations on the Bondi-Sachs metrics. The question is still open for the Einstein field equations. In case of spherically symmetric Bondi-Sachs metrics, the Einstein scalar field equations with required asymptotics can be solved when $\gamma =\delta =U=W = 0$, and $r_0=1$, $r_1=\infty$ \cite{C1, LZ1, LZ2, LZ3}. For asymptotically flat Bondi-Sachs metrics, heuristically, we can solve \eqref{eq:EQ1}-\eqref{eq:EQ7} as follows. On the null hypersurface $\{u=u_0\}$, integrating \eqref{eq:EQ1} with respect to $r$, we obtain
\begin{equation}\label{eq:betaTheory}
    \beta=\int_{r_1}^r \mathscr{R}_1 \mathrm{d}r' +B,
\end{equation}
where
\begin{align*}
B(u_0,\theta,\phi)=\beta(u_0,r_1,\theta,\phi).
\end{align*}
Integrating \eqref{eq:EQ2}, \eqref{eq:EQ3} with respect to $r$, we obtain
\begin{align}
    r^4\mathrm{e}^{-2\beta}\left(\mathrm{e}^{2\gamma}U_r\cosh(2\delta)+W_r\sinh(2\delta) \right)
    &=\int_{r_0}^r\mathscr{R}_2 \mathrm{d}r'-6N :=x , \label{eq:EQ2Integrater} \\
   r^4\mathrm{e}^{-2\beta}
    \left(U_r\sinh(2\delta)+\mathrm{e}^{-2\gamma}W_r\cosh(2\delta)\right) &=\int_{r_0}^r\mathscr{R}_3\mathrm{d}r'-6P:=y, \label{eq:EQ3Integrater}
\end{align}
where
\begin{align*}
    N &=-\frac{1}{6}\left(r^4\mathrm{e}^{-2\beta}\left(\mathrm{e}^{2\gamma}U_r\cosh(2\delta)+W_r\sinh(2\delta) \right)\right)\Big|_{u=u_0, r=r_0}, \\
    P &=-\frac{1}{6}\left(r^4\mathrm{e}^{-2\beta}
    \left(U_r\sinh(2\delta)+\mathrm{e}^{-2\gamma}W_r\cosh(2\delta)\right)\right)\Big|_{u=u_0, r=r_0}
\end{align*}
determined by the initial data $\gamma(u_0)$, $\delta(u_0)$ at $r=r_0$ and the boundary values $\beta(r_0)$, $(r^4U_r)(r_0)$, $(r^4W_r)(r_0)$ at $u=u_0$.
From \eqref{eq:EQ2Integrater}, \eqref{eq:EQ3Integrater}, we can solve $U_r$ and $W_r$, and then integrating them with respect to $r$, we obtain
\begin{align}
    U &=\int_{r_1}^r \mathrm{e}^{2\beta}(r')^{-4}
       \left(\mathrm{e}^{-2\gamma}x\cosh(2\delta) -y \sinh(2\delta)\right) \mathrm{d}r'+ X \label{eq:UTheory}, \\
    W &=\int_{r_1}^r \mathrm{e}^{2\beta}(r')^{-4}
       \left(\mathrm{e}^{2\gamma}y\cosh(2\delta)- x\sinh(2\delta) \right) \mathrm{d}r'+ Y, \label{eq:WTheory}
\end{align}
where
\begin{align*}
X(u_0,\theta,\phi)=U(u_0,r_1,\theta,\phi), \quad Y(u_0,\theta,\phi) =W(u_0,r_1,\theta,\phi).
\end{align*}
Integrating \eqref{eq:EQ4}, we obtain
\begin{equation}\label{eq:VTheory}
    V=\int_{r_0}^r \mathscr{R}_4 \mathrm{d}r'-2M,
\end{equation}
where
\begin{align*}
M(u_0,\theta,\phi)=-\frac{1}{2}V(u_0,r_0,\theta,\phi).
\end{align*}
Note that \eqref{eq:EQ5} and \eqref{eq:EQ6} take the form
\begin{equation}\label{eq:EQ5compress}
    f_r+\zeta \hat{f}=\mathscr{R}_5,
\end{equation}
\begin{equation}\label{eq:EQ6compress}
   \hat{f}_r-\zeta f=\mathscr{R}_6,
\end{equation}
where
\begin{align*}
f =r\gamma_u\cosh(2\delta),  \quad \hat{f} =r\delta_u, \quad \zeta =2\gamma_r\sinh(2\delta).
\end{align*}
Then we can solve $\gamma_u$, $\delta_u$ by integrating \eqref{eq:EQ5compress}, \eqref{eq:EQ6compress} with respect to $r$, as well as $\gamma$, $\delta$ by integrating the expressions of $\gamma_u$, $\delta_u$ with respect to $u$, where integration constants are determined by the boundary values $\gamma_u$, $\delta_u$ at $r=r_0$ and the initial data $\gamma$, $\delta$ at $u=u_0$. Finally, integrating \eqref{eq:EQ7} with respect to $r$ and $u$, we obtain
\begin{equation*}\label{eq:psiuTheory}
    \Psi=\frac{1}{r}\int _{u_0} ^u \int_{r_1}^r\mathscr{R}_7\mathrm{d}r'\mathrm{d}u'   +\frac{1}{r}(r\Psi_u)(u_0, r_1, \theta, \phi) (u-u_0) +\Psi(u_0, r, \theta, \psi).
\end{equation*}
We repeat this procedure to extend $u$ by using the obtained $\gamma(u, r, \theta, \phi)$, $\delta(u, r, \theta, \phi)$ and $\Psi(u, r, \theta, \phi)$ as the new initial data. Then we can get local and global existence and uniqueness. The rigorous and complete proof based on the above procedure needs much more sophisticated estimates on unknowns when $r_0=0$, $r_1=\infty$. We address the proof of this existence and uniqueness elsewhere.

\mysection{Asymptotic Expansions}
\ls

In this section, we study the power series solutions of the Einstein scalar field equations \eqref{eq:EQ1}-\eqref{eq:EQ7} for asymptotically flat Bondi-Sachs metric \eqref{eq:BondiSachsMetric} with asymptotical condition \eqref{eq:BSmetric1}. We follow the idea of \cite{BBM, S, vdB} and assume that $\gamma$, $\delta$ satisfy the outgoing radiating condition
\begin{align}
	\gamma=&\frac{c}{r}+\left(-\frac{1}{6}c^3-\frac{3}{2}d^2c+C\right)\frac{1}{r^3}+O\left(\frac{1}{r^4}\right),\label{eq:gammaExpand} \\
	\delta=&\frac{d}{r}+\left(-\frac{1}{6}d^3+\frac{1}{2}c^2d+D\right)\frac{1}{r^3}+O\left(\frac{1}{r^4}\right).\label{eq:deltaExpand}
\end{align}
The absence of $r^{-2}$ term in the above expansions prevent appearance of $\ln r $ term in the expansions of unknown in vacuum Einstein field equations \cite{BBM, S, vdB}. We refer to \cite{CMS} for polyhomogeneity expansions involving $r^{-2}$ term for $\gamma$ and $\delta$.

\begin{proposition}
Suppose ($\mathscr{L}$, ${\bf g}$) is an asymptotically flat Bondi-Sachs spacetime which satisfies the Einstein scalar field equations. If \eqref{eq:gammaExpand}, \eqref{eq:deltaExpand} hold and scalar field $\Psi$ takes the following expansion
\begin{align*}
\Psi=I(u, \theta, \phi)+\frac{H(u, \theta, \phi)}{r}+ \frac{K(u, \theta, \phi)}{r^2} +  O\left(\frac{1}{r^3}\right),
\end{align*}
then $X(u, \theta, \phi)$, $Y(u, \theta, \phi)$ in \eqref{eq:UTheory}, \eqref{eq:UTheory} must be zero, and $B(u, \theta, \phi)$ in \eqref{eq:betaTheory} can be reduced to zero by suitable coordinate transformations. Moreover, $I(u, \theta, \phi)$ must be constant which can be chosen as zero since the Einstein scalar field equations are invariant for adding any constant to scalar fields.
\end{proposition}
\pf It is straightforward from \eqref{eq:EQ1}-\eqref{eq:EQ4} that, for $r_1=\infty$, 
\begin{align*}
\beta=&B-\frac{2(c^2+d^2)+H^2}{8r^2}+O\left(\frac{1}{r^3}\right),\\ 
U=&X+\frac{2\mathrm{e}^{2B}B_\theta}{r}+O\left(\frac{1}{r^2}\right),\\
W=&Y+\frac{2\mathrm{e}^{2B}B_\phi\csc\theta}{r}+O\left(\frac{1}{r^2}\right),\\
%V=&\bigg(X_\theta +X\cot\theta+Y_\phi\csc\theta \bigg)r^2
%    +\frac{r\mathrm{e}^{2B}}{2} \bigg(2+8B_\theta^2+8B_\phi^2\csc^2\theta\\
%  &+4B_{\theta\theta}+4B_\theta\cot\theta+4B_{\phi\phi}\csc^2\theta -I_\theta^2-I_\phi^2\csc^2\theta \bigg)-2M\\
%  &+O\left(\ln r\right)+O\left(\frac{1}{r}\right).
    V=& \bigg(X_\theta +X\cot\theta+Y_\phi\csc\theta \bigg)r^2+O(r)
\end{align*}
Thus,
\begin{align*}
g_{00}=&\bigg(X^2+Y^2\bigg)r^2+r\bigg( 4dXY-\mathrm{e}^{2B}X\cot\theta+4\mathrm{e}^{2B}X B_\theta\\
       &+4\mathrm{e}^{2B} Y B_\phi \csc\theta -\mathrm{e}^{2B}X_\theta -\mathrm{e}^{2B}Y_\phi\csc\theta+2cX^2-2c Y^2 \bigg)+O(1).
\end{align*}
Using \eqref{eq:BSmetric1}, we obtain
\begin{align*}
X=Y=0.
\end{align*}

Next we show that, similar to \cite{BBM}, we can reduce $B=0$ by a coordinate transformation
\begin{equation}\label{eq:coordinateTransform}
    \left\{ \begin{aligned}
        u &=z_0(\hat{u},\hat{\theta},\hat{\phi}),
        +\frac{u_1(\hat{u},\hat{\theta},\hat{\phi})}{\hat{r}}+O\left(\frac{1}{\hat{r}^2}\right), \\
        r &=\hat{r}+r_0(\hat{u},\hat{\theta},\hat{\phi})+O\left(\frac{1}{\hat{r}}\right), \\
        \theta &=\hat{\theta}+\frac{\theta_1(\hat{u},\hat{\theta},\hat{\phi})}{\hat{r}}+O\left(\frac{1}{\hat{r}^2}\right), \\
        \phi &=\hat{\phi}+\frac{\phi_1(\hat{u},\hat{\theta},\hat{\phi})}{\hat{r}}+O\left(\frac{1}{\hat{r}^2}\right).
    \end{aligned} \right.
\end{equation}
Denote $(\hat{x}^\alpha)=(\hat{u},\hat{r},\hat{\theta},\hat{\phi})$ and
\begin{align*}
\hat{g}_{\mu\nu}={\bf g}\left(\frac{\partial}{\partial \hat{x}^\mu},\frac{\partial}{\partial \hat{x}^\nu}\right).
\end{align*}
Using \eqref{eq:coordinateTransform}, we obtain
\begin{align*}
\hat{g}_{01}
=&-(z_0)_{\hat{u}}(\hat{u},\hat{\theta},\hat{\phi})\mathrm{e}^{2B\left(z_0(\hat{u},\hat{\theta},\hat{\phi}),\hat{\theta},\hat{\phi}\right)}
    +O\left(\frac{1}{\hat{r}}\right),
%\label{eq:newg01}
\\
\hat{g}_{11} =&\frac{1}{\hat{r}^2}\bigg(2u_1(\hat{u},\hat{\theta},\hat{\phi})\mathrm{e}^{2B\left(z_0(\hat{u},\hat{\theta},\hat{\phi}),\hat{\theta},\hat{\phi}\right)}
+\theta_1 ^2 (\hat{u},\hat{\theta},\hat{\phi})\\
&+\phi_1 ^2(\hat{u},\hat{\theta},\hat{\phi}) \sin^2\hat{\theta} \bigg)
    +O\left(\frac{1}{\hat{r}^3}\right),
%\label{eq:newg11}
\\
\hat{g}_{12}
=&-(z_0)_{\hat{\theta}}(\hat{u},\hat{\theta},\hat{\phi})\mathrm{e}^{2 B\left( z_0(\hat{u},\hat{\theta},\hat{\phi}),\hat{\theta},\hat{\phi}\right)}
    -\theta_1(\hat{u},\hat{\theta},\hat{\phi}) +O\left(\frac{1}{\hat{r}}\right),
%\label{eq:newg12}
\\
\hat{g}_{13}
=&-(z_0)_{\hat{\phi}}(\hat{u},\hat{\theta},\hat{\phi})\mathrm{e}^{2 B\left( z_0(\hat{u},\hat{\theta},\hat{\phi}),\hat{\theta},\hat{\phi}\right)}
-\phi_1(\hat{u},\hat{\theta},\hat{\phi})\sin^2\hat{\theta} +O\left(\frac{1}{\hat{r}}\right).
%\label{eq:newg13}
\end{align*}
Therefore,
\begin{align*}
%\label{eq:newg2233minus23sq}
\hat{g}_{22}&\hat{g}_{33}-\hat{g}_{23}^2-\hat{r}^4\sin^2\hat{\theta} \\
=&2\hat{r}^3 \bigg(-2(z_0)_{\hat{\phi}}(\hat{u},\hat{\theta},\hat{\phi})
\mathrm{e}^{2B(z_0(\hat{u},\hat{\theta},\hat{\phi}),\hat{\theta},\hat{\phi})}B_\phi(z_0(\hat{u},\hat{\theta},\hat{\phi}),\hat{\theta},\hat{\phi})
\csc^2\theta \\
&-2(z_0)_{\hat{\theta}}(\hat{u},\hat{\theta},\hat{\phi})\mathrm{e}^{2B(z_0(\hat{u},\hat{\theta},\hat{\phi}),\hat{\theta},\hat{\phi})}
B_\theta(z_0(\hat{u},\hat{\theta},\hat{\phi}),\hat{\theta},\hat{\phi}) \\
&+(\phi_1)_{\hat{\phi}}(\hat{u},\hat{\theta},\hat{\phi})+(\theta_1)_{\hat{\theta}}(\hat{u},\hat{\theta},\hat{\phi})
    +\theta_1(\hat{u},\hat{\theta},\hat{\phi})\cot\hat{\theta} \\
&+2r_0(\hat{u},\hat{\theta},\hat{\phi})\bigg) \sin^2\theta +O\left(\hat{r}^2\right).
\end{align*}
Let $z_0$ satisfy
\begin{equation*}
%\label{eq:u0fromB}
    (z_0)_{\hat{u}}(\hat{u},\hat{\theta},\hat{\phi})=\mathrm{e}^{-2B\left( z_0(\hat{u},\hat{\theta},\hat{\phi}),\hat{\theta},\hat{\phi}\right)}.
\end{equation*}
We obtain
\begin{align*}
\hat{B}=0.
\end{align*}

With $B=X=Y=0$, the left hand sides of \eqref{eq:EQ5}, \eqref{eq:EQ6} are both $O\left(\frac{1}{r^3}\right)$, but 
\begin{align*}
\mathscr{R}_5=&\frac{I_\theta^2-I_\phi^2\csc^2\theta}{4r} +O\left(\frac{1}{r^2}\right),\\
\mathscr{R}_6=&\frac{I_\theta I_\phi \csc\theta}{2r}+O\left(\frac{1}{r^2}\right).
\end{align*}
Therefore, \eqref{eq:EQ5}, \eqref{eq:EQ6} give 
\begin{align*}
I_\theta^2-I_\phi^2\csc^2\theta=I_\theta I_\phi \csc\theta=0 \Longrightarrow I _\theta =I _\phi =0.
\end{align*}
With this condition, 
\begin{align*}
\mathscr{R}_7 =O\left(\frac{1}{r^2}\right).
\end{align*}
Therefore, \eqref{eq:EQ7} gives 
\begin{align*}
I_u -\frac{K_u}{r^2} =O\left(\frac{1}{r^2}\right) \Longrightarrow I_u =0.
\end{align*}
So we can choose $I$ to be zero. \qed

From now on, take the following power series expansion
\begin{equation}\label{eq:psiExpand}
    \Psi=\frac{H}{r}+\frac{K}{r^2}+\frac{L}{r^3}+O\left(\frac{1}{r^4}\right).
\end{equation}
Substituting \eqref{eq:gammaExpand}, \eqref{eq:deltaExpand}, \eqref{eq:psiExpand} into \eqref{eq:betaTheory}, \eqref{eq:EQ2Integrater}, \eqref{eq:EQ3Integrater}, \eqref{eq:UTheory} and \eqref{eq:WTheory}, we obtain the following asymptotic expansions:
\begin{align}
    \beta =&-\frac{2(c^2+d^2)+H^2}{8r^2} -\frac{HK}{3r^3} +\frac{\overset{4}{\beta}}{r^4}+O\left(\frac{1}{r^5}\right), \label{eq:betaExpand}\\
    U =&-\frac{l}{r^2}+\frac{2( 2cl+2d\hat{l}+3N)}{3r^3}+ \frac{\overset{4}{U}}{r^4}+O\left(\frac{1}{r^5}\right), \label{eq:UExpand} \\
    W =&-\frac{\hat{l}}{r^2} +\frac{2(2dl-2c\hat{l}+3P)}{3r^3} +\frac{\overset{4}{W}}{r^4} +O\left(\frac{1}{r^5}\right), \label{eq:WExpand} \\
    V =& r-2\mathcal{M}-\big(l_\theta+l\cot\theta+\hat{l}_\phi\csc\theta\big)+\frac{\overset{1}{V}}{r}+O\left(\frac{1}{r^2}\right),  \label{eq:VExpand}
\end{align}
where $l$, $\hat{l}$ are given by \eqref{l-barl}. Using \eqref{eq:EQ5}, \eqref{eq:EQ6}, we obtain
\begin{align}
    \gamma_u=&\frac{c_u}{r}+\frac{\overset{3}{\gamma} _u}{r^3}+O\left(\frac{1}{r^4}\right),\label{gamma-u}\\
    \delta_u=&\frac{d_u}{r}+\frac{\overset{3}{\delta} _u}{r^3}+O\left(\frac{1}{r^4}\right).\label{delta-u}
\end{align}
Substituting \eqref{eq:gammaExpand} into \eqref{gamma-u} and \eqref{eq:deltaExpand} into \eqref{delta-u}, we obtain
\begin{align}
    -\frac{c^2c_u}{2}-3d d_u c-\frac{3d^2 c_u}{2}+C_u &=\overset{3}{\gamma} _u, \label{eq:Cdetermined}\\
    -\frac{d^2 d_u}{2}+c c_u d+\frac{c^2 d_u}{2}+D_u &=\overset{3}{\delta} _u. \label{eq:Ddetermined}
\end{align}
Substituting \eqref{eq:Cdetermined}, \eqref{eq:Ddetermined} into \eqref{eq:EQ7}, we obtain
\begin{equation}\label{eq:psiuExpand}
\Psi_u=\frac{H_u}{r}-\frac{H_{\theta\theta}+H_\theta\cot\theta+H_{\phi\phi}\csc^2\theta}{2r^2}+\frac{\overset{3}{\Psi} _u}{r^3}
    +O\left(\frac{1}{r^4}\right).
\end{equation}
We refer to Appendix B for explicit expressions of $\overset{4}{\beta}$, $\overset{4}{U}$, $\overset{4}{W}$, $\overset{1}{V}$, $\overset{3}{\gamma_u}$, $\overset{3}{\delta} _u$, $\overset{3}{\Psi}$.

Substituting \eqref{eq:psiExpand} into \eqref{eq:psiuExpand}, we obtain
\begin{align*}
2 K_u=&-H_{\theta\theta}-H_\theta\cot\theta-H_{\phi\phi}\csc^2\theta,\\
4 L_u=& 4c_\theta H_\theta-4 c_\phi H_\phi\csc^2\theta+2Hc_{\theta\theta}+6 Hc_\theta\cot\theta \\
     & -2 Hc_{\phi\phi}\csc^2\theta +2c\big(H_{\theta\theta}+3H_\theta\cot\theta -H_{\phi\phi}\csc^2\theta\\
     & -2H\big)+4 d_\theta H_\phi\csc\theta +4 d_\phi H_\theta\csc\theta +4 Hd_{\theta\phi}\csc\theta \\
     & +4H d_\phi\cot\theta\csc\theta +4 dH_{\theta\phi}\csc\theta +4 dH_{\phi}\cot\theta\csc\theta \\
     & +2 H \mathcal{M} - K_{\theta\theta}-K_\theta \cot\theta- K_{\phi\phi}\csc^2\theta -2 K.
\end{align*}
Finally, comparing the coefficients of $r^{-2}$ term in three supplementary equations \eqref{eq:supplementary3eq}, we obtain
\begin{align}
    \mathcal{M}_u=&-c_u^2-d_u^2-\frac{H_u^2}{2},\label{eq:R00eqExpand}
\end{align}    
and
\begin{align*} 
    6N_u =&4c_u d_\phi\csc\theta+4d_\phi\csc^3\theta -4d_u c_\phi\csc\theta+4d c_{u\phi}\csc\theta       \\
          &+2c_\theta-c_{\theta\theta\theta}+3c_uc_\theta-cc_{u\theta}-3c_{\theta\theta}\cot\theta+3c_\theta\csc^2\theta    \\
          &+3c_{\theta\phi\phi}\csc^2\theta-4c d_{u\phi}\csc\theta +3d_u d_\theta-d d_{u\theta}     \\
          %\label{eq:R02eqExpand}   \\
          &+d_{\phi\phi\phi}\csc^3\theta-3d_{\theta\theta\phi}\csc\theta-3d_{\theta\phi}\cot\theta\csc\theta  \\
          &+\frac{3 H_u H_\theta}{2}-\frac{H H_{u\theta}}{2}-2\mathcal{M}_\theta, \\
    6P_u =&-4c_u d_\theta+4c_\theta d_u-4dc_{u\theta}-8dc_u\cot\theta-d d_{u\phi}\csc\theta   \\
          &-3c_{\theta\theta\phi}\csc\theta -cc_{u\phi}\csc\theta-3c_{\theta\phi}\cot\theta\csc\theta    \\
          &+c_{\phi\phi\phi}\csc^3\theta+3c_uc_\phi\csc\theta +4c_\phi\csc^3\theta +4cd_{u\theta}  \\
          &+8 cd_u\cot\theta+3d_u d_\phi\csc\theta-d_\theta +d_{\theta\theta\theta} +3 d_{\theta\theta}\cot\theta \\
          %\label{eq:R03eqExpand}\\
          &-4d_\theta\csc^2\theta-3d_{\theta\phi\phi}\csc^2\theta+d_\theta \cot^2\theta  \\
          &+\frac{3H_u H_\phi}{2}\csc\theta-\frac{H H_{u\phi}}{2} \csc\theta-2\mathcal{M}_\phi\csc\theta. 
\end{align*}

Inspired by \eqref{eq:R00eqExpand}, we can provide the following definition.
\begin{definition}\label{modified mass aspect}
$c_u$, $d_u$ and $H_u$ are defined as {\it news functions}, and $\mathcal{M}$ is defined as {\it mass aspect} of the Einstein scalar field equations for \eqref{eq:BondiSachsMetric}.
\end{definition}

\mysection{Peeling Property}
\ls

In this section, we prove the peeling property for the Einstein scalar fields when spacetimes are asymptotically flat Bondi-Sachs and the scalar field $\Psi$ satisfies \eqref{eq:psiExpand}.

Peeling property indicates that the Weyl curvatures of asymptotically flat spacetimes have good asymptotic behaviors, which plays a significant role for extracting gravitational waves from numerical simulation. When spacetimes are vacuum and asymptotically simple, Newman and Penrose introduced a complex null tetrad and proved this property \cite{NP1, NP2, NU}. This procedure is referred to as the Newman-Penrose formalism. For vacuum Bondi-Sachs metrics, peeling property holds also when the cosmological constant is zero \cite{BBM, S} or nonzero \cite{XZ}. In \cite{HC2}, He and Cao proved the peeling property for Einstein scalar field equations with zero cosmological constant via Newman-Penrose formalism.  As it is unclear whether Newman-Penrose and Bondi-Sachs coordinates are equivalent, the peeling property is also important via Bondi-Sachs formalism.

Denote $\mathrm{i}=\sqrt{-1}$. The Bondi-Sachs metric \eqref{eq:BondiSachsMetric} has the following null tetrad \cite{XZ}
\begin{equation}\label{eq:NPbasis}
\begin{aligned}
    \hat{e}_0 &=\boldsymbol{l}=\mathrm{e}^{-2\beta}\left(\frac{\partial}{\partial u}-\frac{V}{2r}\frac{\partial}{\partial r}+U\frac{\partial}{\partial\theta}+W\csc\theta \frac{\partial}{\partial \phi}\right), \\
    \hat{e}_1 &=\boldsymbol{k} =\frac{\partial}{\partial r}, \\
    \hat{e}_2 &=\boldsymbol{m} =\frac{\mathrm{e}^{-\gamma}\big(1-\mathrm{i}\sinh(2\delta)\big)}{r\sqrt{2\cosh(2\delta)}}\frac{\partial}{\partial \theta}
    +\frac{\mathrm{i}\mathrm{e}^\gamma \sqrt{\cosh(2\delta)}\csc\theta}{\sqrt{2}r}\frac{\partial}{\partial \phi}, \\
    \hat{e}_3 &=\bar{\boldsymbol{m}}=\frac{\mathrm{e}^{-\gamma}\big(1+\mathrm{i}\sinh(2\delta)\big)}{r\sqrt{2\cosh(2\delta)}}\frac{\partial}{\partial\theta}
    -\frac{\mathrm{i}\mathrm{e}^\gamma\sqrt{\cosh(2\delta)}\csc\theta}{\sqrt{2}r}\frac{\partial}{\partial \phi}.
\end{aligned}
\end{equation}
Under this null tetrad, both the metric matrix $(\hat{g}_{\mu\nu})=({\bf g}(\hat{e}_\mu,\hat{e}_\nu))$ and its inverse matrix $(\hat{g}^{\mu\nu})$ are
\begin{equation*}
    \begin{pmatrix}
        0 & -1 & 0 & 0 \\
        -1 & 0 & 0 & 0 \\
        0 & 0 & 0 & 1 \\
        0 & 0 & 1 & 0
    \end{pmatrix}.
\end{equation*}
The structure coefficients $\hat{C}_{\mu\nu}^\sigma$ of the tetrad satisfy
\begin{equation*}
%\label{eq:StructureConstantDef}
   [\hat{e}_\mu,\hat{e}_\nu]=\hat{C}_{\mu\nu}^\sigma \hat{e}_\sigma.
\end{equation*}
Denote
\begin{align*}
\hat{C}_{\mu\nu\sigma}=\hat{C}_{\mu\nu}^\tau \hat{g}_{\tau\sigma}.
\end{align*}
Use the Koszul formula \cite{O0}
\begin{align*}
	2\langle \nabla_X Y,Z\rangle
    &=X\langle Y,Z\rangle+Y\langle Z,X\rangle-Z\langle X,Y\rangle \\
	&\quad +\langle [X,Y],Z\rangle -\langle [Y,Z],X\rangle
	+\langle [Z,X],Y\rangle,
\end{align*}
where $\langle X,Y\rangle=g(X,Y)$, we obtain connection coefficients
\begin{equation*}
%\label{eq:NPConnection3down}
\hat{\Gamma}_{\mu\nu\sigma}=\langle \nabla_{\hat{e}_\mu}\hat{e}_\nu,\hat{e}_\sigma\rangle
=\frac{1}{2}\left(\hat{C}_{\mu\nu\sigma}-\hat{C}_{\nu\sigma\mu}+\hat{C}_{\sigma\mu\nu}\right).
\end{equation*}

We introduce the spin coefficients defined in \cite{O}.
\begin{definition}
%\label{def:SpinCoefficientDef}
The spin coefficients of the null tetrad \eqref{eq:NPbasis} are twelve complex-valued functions
\begin{align*}
%\label{eq:SpinCoefficientDefGroup1}
    \boldsymbol{\kappa} &=-\langle \nabla_{\boldsymbol{k}} \boldsymbol{k},\boldsymbol{m}\rangle=-\hat{\Gamma}_{112}, \\
    \boldsymbol{\rho} &=-\langle \nabla_{\bar{\boldsymbol{m}}}\boldsymbol{k},\boldsymbol{m}\rangle=-\hat{\Gamma}_{312}, \\
    \boldsymbol{\sigma} &=-\langle \nabla_{\boldsymbol{m}} \boldsymbol{k},\boldsymbol{m}\rangle=-\hat{\Gamma}_{212}, \\
    \boldsymbol{\tau} &=-\langle \nabla_{\boldsymbol{l}} \boldsymbol{k},\boldsymbol{m}\rangle=-\hat{\Gamma}_{012},\\
%\label{eq:SpinCoefficientDefGroup2}
    \boldsymbol{\nu} &=\langle \nabla_{\boldsymbol{l}} \boldsymbol{l},\bar{\boldsymbol{m}}\rangle=\hat{\Gamma}_{003}, \\
    \boldsymbol{\mu} &=\langle\nabla_{\boldsymbol{m}} \boldsymbol{l},\bar{\boldsymbol{m}}\rangle=\hat{\Gamma}_{203}, \\
    \boldsymbol{\lambda} &=\langle\nabla_{\bar{\boldsymbol{m}}}\boldsymbol{l},\bar{\boldsymbol{m}}\rangle=\hat{\Gamma}_{303}, \\
    \boldsymbol{\pi} &=\langle\nabla_{\boldsymbol{k}} \boldsymbol{l},\bar{\boldsymbol{m}}\rangle=\hat{\Gamma}_{103},\\
%\label{eq:SpinCoefficientDefGroup3}
    \boldsymbol{\varepsilon} &=\frac{1}{2}
    \Big(-\langle \nabla_{\boldsymbol{k}}\boldsymbol{k},\boldsymbol{l}\rangle+\langle \nabla_{\boldsymbol{k}} \boldsymbol{m},\bar{\boldsymbol{m}}\rangle\Big)
=\frac{1}{2}\Big(-\hat{\Gamma}_{110}+\hat{\Gamma}_{123}\Big), \\
   \boldsymbol{\gamma} &=\frac{1}{2}\Big(\langle\nabla_{\boldsymbol{l}} \boldsymbol{l},\boldsymbol{k}\rangle-\langle\nabla_{\boldsymbol{l}}\bar{\boldsymbol{m}},\boldsymbol{m}\rangle \Big)=\frac{1}{2}\left(\hat{\Gamma}_{001}-\hat{\Gamma}_{032}\right), \\
   \boldsymbol{\beta} &=\frac{1}{2}
    \Big(-\langle \nabla_{\boldsymbol{m}}\boldsymbol{k},\boldsymbol{l}\rangle+\langle\nabla_{\boldsymbol{m}} \boldsymbol{m},\bar{\boldsymbol{m}}\rangle\Big)
=\frac{1}{2}
\left(-\hat{\Gamma}_{210}+\hat{\Gamma}_{223}\right), \\
    \boldsymbol{\alpha} &=\frac{1}{2}\Big(\langle \nabla_{\bar{\boldsymbol{m}}}\boldsymbol{l},\boldsymbol{k}\rangle-\langle \nabla_{\bar{\boldsymbol{m}}}\bar{\boldsymbol{m}},\boldsymbol{m}\rangle\Big)=\frac{1}{2}\left(\hat{\Gamma}_{301}-\hat{\Gamma}_{332}\right).
\end{align*}
\end{definition}

It is straightforward that the spin coefficients of Bondi-Sachs metric \eqref{eq:BondiSachsMetric}
\begin{align*}
%\label{eq:BSkapparhoExpression}
\boldsymbol{\kappa} =&0,\\
\boldsymbol{\rho}   =&-\frac{1}{r},\\
\boldsymbol{\sigma} =&-\gamma_r-\tanh(2\delta)\delta_r+\mathrm{i}\bigg(\sinh(2\delta)\gamma_r-\operatorname{sech}(2\delta)\delta_r\bigg),\\
\boldsymbol{\tau}   =&\frac{1}{2\sqrt{2}r\sqrt{\cosh(2\delta)}}
    \bigg( -2 \mathrm{e}^{-\gamma}\beta_\theta \\
    &+\mathrm{e}^{-2\beta +\gamma}r^2 U_r\cosh(2\delta)+\mathrm{e}^{-2\beta -\gamma}r^2 W_r\sinh(2\delta)\\  &+\mathrm{i}\Big( -2\mathrm{e}^\gamma\beta_\phi\cosh(2\delta)\csc\theta
    +2\mathrm{e}^{-\gamma}\beta_\theta\sinh(2\delta)+\mathrm{e}^{-2\beta-\gamma}r^2 W_r\Big) \bigg),  \\
\boldsymbol{\nu} =&\frac{1}{2\sqrt{2} r^2\sqrt{\cosh(2\delta)} }\bigg(\mathrm{e} ^{-2\beta-\gamma} V_\theta\\
    &+\mathrm{i}\Big(\mathrm{e}^{-2\beta-\gamma}V_\theta\sinh(2\delta)-\mathrm{e}^{-2\beta+\gamma}V_\phi \csc\theta \cosh(2\delta) \Big)\bigg),  \\ 
\boldsymbol{\mu} =&\frac{\mathrm{e}^{-2\beta}}{2}\bigg(U\cot\theta+ U_\theta+W_\phi\csc\theta -\frac{V}{r^2} \bigg),  \\
\boldsymbol{\lambda} =&\frac{\mathrm{e}^{-2\beta}}{8}
    \bigg(8  \mathrm{e}^{-2\gamma} \tanh(2\delta) (W_\theta - W\cot\theta)
     -\frac{4V}{r}(\delta_r\tanh(2\delta)+\gamma_r) \\
    &  -4 U(\cot\theta-2\delta_\theta \tanh(2\delta)-2\gamma_\theta)
      +8 W\csc\theta ( \delta_\phi \tanh(2\delta)+\gamma_\phi) \\
	& - 4 (W_\phi\csc\theta-U_\theta-2\delta_u \tanh(2\delta)-2\gamma_u ) \\
    &-4\mathrm{i} \Big( \cosh(2\delta)  \mathrm{e}^{2\gamma}U_\phi \csc\theta 
                      -\left(2\operatorname{sech}(2\delta)-\cosh(2\delta)\right)\mathrm{e}^{-2\gamma}( W\cot\theta\\
    &-W_\theta) +\operatorname{sech}(2\delta) \Big(2U \delta _\theta +2W \delta _\phi \csc\theta  -\frac{V}{r} \delta_r +2\delta _u \Big)
     +\sinh(2\delta) \\
    & \Big(U \cot\theta
     -2 U \gamma_\theta -U_\theta -2W \gamma _\phi \csc\theta +W _\phi \csc\theta +\frac{V}{r}\gamma _r -2\gamma_u\Big)\Big)\bigg),\\      
\boldsymbol{\pi} =&\frac{\mathrm{e}^{-2\beta-\gamma}}{2\sqrt{2}r \sqrt{\cosh(2\delta)}}
    \bigg( 2\mathrm{e}^{2\beta}\beta_\theta+\mathrm{e}^{2\gamma}r^2 U_r \cosh(2\delta) +r^2 W_r\sinh(2\delta)\\
    & -\mathrm{i}\Big( 2\mathrm{e}^{2\gamma+2\beta}\beta_\phi\csc\theta \cosh(2\delta)
      -2\mathrm{e}^{2\beta}\beta_\theta \sinh(2\delta)+r^2 W_r\Big)\bigg),\\
\boldsymbol{\varepsilon} =&\beta_r+\frac{\mathrm{i}\operatorname{sech}(2\delta)}{4}\bigg( \sinh(4\delta)\gamma_r-2\delta_r \bigg),  \\
\boldsymbol{\gamma} =&\frac{\mathrm{e}^{-2\beta}}{8}\bigg( -\frac{2V}{r^2}+\frac{2V_r}{r}  \\
    &-2\mathrm{i} \Big( \cosh(2\delta)  \Big(  \mathrm{e}^{2\gamma} U_\phi \csc\theta  + \mathrm{e}^{-2\gamma}(W\cot\theta-W_\theta)\Big)\\
    &+\operatorname{sech}(2\delta) \Big(2U \delta _\theta +2W \delta _\phi \csc\theta   -\frac{V}{r} \delta_r +2\delta _u \Big)
     +\sinh(2\delta) \\
    & \Big(U \cot\theta
     -2 U \gamma_\theta -U_\theta -2W \gamma _\phi \csc\theta +W _\phi \csc\theta +\frac{V}{r}\gamma _r -2\gamma_u\Big)\Big)\bigg),\\
\boldsymbol{\beta} =&\frac{\mathrm{e}^{-\gamma}}{4\sqrt{2}r \sqrt{\cosh(2\delta)}}\bigg( 2\beta_\theta
	-2\gamma_\theta +2\cot\theta -2\delta_\theta \tanh(2\delta)\\
    & +r^2 \mathrm{e}^{2\gamma-2\beta}U_r \cosh(2\delta)+\mathrm{e}^{-2\beta}r^2 W_r\sinh(2\delta)\\
    &+2\mathrm{i}\Big( \cosh(2\delta) \Big(\mathrm{e}^{2\gamma}\beta_\phi\csc\theta +\mathrm{e}^{2\gamma}\gamma _\phi\csc\theta -2\delta _\theta \Big)
    +\frac{\mathrm{e}^{-2\beta}}{2}r^2 W_r \\
    &+\delta_\theta \tanh(2 \delta) \sinh(2\delta)+\sinh(2\delta)\Big( \mathrm{e}^{2\gamma}\delta_\phi\csc\theta -\beta_\theta +\gamma_\theta -\cot\theta \Big)\bigg),\\
\boldsymbol{\alpha} =&\frac{\mathrm{e}^{-\gamma}}{4\sqrt{2}r \sqrt{\cosh(2\delta)}}\bigg( 2\beta_\theta
	+2\gamma_\theta -2\cot\theta +2\delta_\theta \tanh(2\delta)\\
    & +r^2 \mathrm{e}^{2\gamma-2\beta}U_r \cosh(2\delta)+\mathrm{e}^{-2\beta}r^2 W_r\sinh(2\delta)\\
    &+2\mathrm{i}\Big( \cosh(2\delta) \Big(-\mathrm{e}^{2\gamma}\beta_\phi\csc\theta +\mathrm{e}^{2\gamma}\gamma _\phi\csc\theta -2\delta _\theta \Big)
    -\frac{\mathrm{e}^{-2\beta}}{2}r^2 W_r \\
    &+\delta_\theta \tanh(2 \delta) \sinh(2\delta)+\sinh(2\delta)\Big( \mathrm{e}^{2\gamma}\delta_\phi\csc\theta +\beta_\theta +\gamma_\theta -\cot\theta \Big)\bigg).
\end{align*}
It is clear that
\begin{equation*}
%\label{eq:BSsmallrelation}
    \bar{\boldsymbol{\mu}}=\boldsymbol{\mu}, \quad \boldsymbol{\alpha}+\bar{\boldsymbol{\beta}}=\boldsymbol{\pi}.
\end{equation*}

Denote by $\mathcal{W}$ the $(0,4)$ type Weyl tensor. The Weyl scalars are \cite{O}
\begin{align*}
    \Phi _0 =\mathcal{W}_{\boldsymbol{k}\boldsymbol{m}\boldsymbol{k}\boldsymbol{m}}, \,\,
    \Phi_1 =\mathcal{W}_{\boldsymbol{k}\boldsymbol{l}\boldsymbol{k}\boldsymbol{m}}, \,\,
    \Phi_2 =\mathcal{W}_{\boldsymbol{k}\boldsymbol{m}\bar{\boldsymbol{m}}\boldsymbol{l}}, \,\,
    \Phi_3 =\mathcal{W}_{\boldsymbol{l}\boldsymbol{k}\boldsymbol{l}\bar{\boldsymbol{m}}}, \,\,
    \Phi_4 =\mathcal{W}_{\boldsymbol{l}\bar{\boldsymbol{m}}\boldsymbol{l}\bar{\boldsymbol{m}}}.
\end{align*}

\begin{theorem}
Suppose ($\mathscr{L}$, ${\bf g}$) is an asymptotically flat Bondi-Sachs spacetime which satisfies the Einstein scalar field equations. If \eqref{eq:gammaExpand}, \eqref{eq:deltaExpand} hold and scalar field $\Psi$ takes expansion \eqref{eq:psiExpand}, then the peeling property
\begin{align*}
\Phi_k=\frac{f_k}{r^{5-k}}+O\left(\frac{1}{r^{6-k}}\right) ,\quad k=0,\ldots,4,
\end{align*}
holds, where $f_k$ are functions depending only on $u$, $\theta$, $\phi$.
\end{theorem}
\pf It is straightforward that
\begin{align*}
\Phi_0 = & \boldsymbol{k}(\boldsymbol{\sigma})
    -\boldsymbol{\sigma}(2\boldsymbol{\rho}+3\boldsymbol{\varepsilon}-\bar{\boldsymbol{\varepsilon}}),  \\
\Phi_1 = & \boldsymbol{k}(\boldsymbol{\beta})-\boldsymbol{m}(\boldsymbol{\varepsilon})
    -(\boldsymbol{\alpha}+\boldsymbol{\pi})\boldsymbol{\sigma}
    -(\boldsymbol{\rho}+\boldsymbol{\varepsilon}-\bar{\boldsymbol{\varepsilon}})\boldsymbol{\beta}, \\
\Phi_2 =& \boldsymbol{k}(\boldsymbol{\mu})-\boldsymbol{m}(\boldsymbol{\pi})
    +\frac{\boldsymbol{l}(\Psi)\boldsymbol{k}(\Psi)}{6}
    -\frac{\boldsymbol{m}(\Psi)\bar{\boldsymbol{m}}(\Psi)}{6}-\boldsymbol{\rho}\boldsymbol{\mu}\\
    &-\boldsymbol{\sigma}\boldsymbol{\lambda}-2\boldsymbol{\beta}\boldsymbol{\pi}
    +(\boldsymbol{\varepsilon}+\bar{\boldsymbol{\varepsilon}})\boldsymbol{\mu},  \\
%\label{eq:NPPsiBondiSachsFormula}
\Phi_3 =&\boldsymbol{k}(\boldsymbol{\nu})-\boldsymbol{l}(\boldsymbol{\pi})
    -\boldsymbol{\mu}(\boldsymbol{\pi}+\bar{\boldsymbol{\tau}})
    -\boldsymbol{\lambda}(\bar{\boldsymbol{\pi}}+\boldsymbol{\tau})
    -\boldsymbol{\pi}(\boldsymbol{\gamma}-\bar{\boldsymbol{\gamma}}) \\
    &+\boldsymbol{\nu}(3\boldsymbol{\varepsilon}+\bar{\boldsymbol{\varepsilon}})
     -\frac{1}{2}\boldsymbol{l}(\Psi)\bar{\boldsymbol{m}}(\Psi),\\
\Phi_4 =&-\boldsymbol{l}(\boldsymbol{\lambda})+\bar{\boldsymbol{m}}(\boldsymbol{\nu})
         -(2\boldsymbol{\mu}+3\boldsymbol{\gamma}-\bar{\boldsymbol{\gamma}})\boldsymbol{\lambda}
         +(2\boldsymbol{\alpha}+2\boldsymbol{\pi}-\bar{\tau})\boldsymbol{\nu}.
\end{align*}
Using the asymptotic expansions, we obtain
\begin{align*}
    \Phi_0 =&-\frac{12C+cH^2+\mathrm{i}(12 D+dH^2)}{2r^5}+O\left(\frac{1}{r^6}\right), \\
    \Phi_1 =&\frac{1}{4\sqrt{2}r^4}\bigg( 4dc_\phi\csc\theta-10cc_\theta-4cd_\phi\csc\theta \\
            &-8c^2\cot\theta-10dd_\theta -8 d^2\cot\theta +12N-HH_\theta \\
            &-\mathrm{i}\Big(4dc_\theta+10 cc_\phi\csc\theta-4c d_\theta+10dd_\phi\csc\theta\\
            &-12P+HH_\phi\csc\theta\Big)\bigg)+O\left(\frac{1}{r^5}\right),\\
    \Phi_2 =&\frac{1}{6r^3}\bigg(-6cc_u-6dd_u-6\mathcal{M} + HH_u +6c -9 c_\theta\cot\theta \\
            &-3c_{\theta\theta}+3 c_{\phi\phi}\csc^2\theta-6d_{\theta\phi}\csc\theta-6 d_\phi\cot\theta\csc\theta \\
            &-\mathrm{i}\Big(6 dc_u +6 c_{\theta\phi}\csc\theta +6c_\phi \cot\theta\csc\theta -6cd_u -3d_{\theta\theta}\\
            &-9 d_\theta\cot\theta+3d_{\phi\phi}\csc^2\theta+6 d\Big)\bigg)+O\left(\frac{1}{r^4}\right),\\
    \Phi_3 =&-\frac{l_u-\mathrm{i}\hat{l}_u}{\sqrt{2}r^2}+O\left(\frac{1}{r^3}\right), \\
    \Phi_4 =&-\frac{c_{uu}-\mathrm{i}d_{uu}}{r}+O\left(\frac{1}{r^2}\right).
\end{align*}
Therefore the theorem follows. \qed

\mysection{Asymptotically Null Spacelike Hypersurfaces}
\ls

In this section, we study the geometry of asymptotically null spacelike hypersurfaces. As metrics of null hypersurfaces degenerate and many geometric properties are lost, asymptotically null spacelike hypersurfaces play an important role in understanding null infinity.

In Minkowski spacetime, the hypersurface
\begin{align*}
u=\sqrt{1+r ^2} -r
\end{align*}
is hyperbola $\mathbb{H}^3$ equipped with the standard hyperbolic metric $\breve{g}$. Let $\{\breve{e} _i \}$ be the frame
 \begin{align*}
\breve{e} _1 = \sqrt{1+r ^2} \frac{\pa}{\partial r}, \quad \breve{e} _2 = \frac{1}{r} \frac{\pa}{\partial \theta}, \quad
\breve{e} _3 = \frac{1}{r \sin \theta} \frac{\pa}{\partial \phi}.
 \end{align*}
Let $\{\breve{e} ^i\}$ be the coframe. Denote $\breve{\nabla }_{i} =\breve{\nabla } _{\breve{e} _i}$, where $\breve{\nabla} $ is the Levi-Civita connection of $\breve{g}$. The connection 1-forms $\{\breve{\omega} _{ij}\}$ are given by
\begin{align*}
\breve{\nabla}_{\breve{e}_i}\breve{e}_j=\tensor{\breve{\omega}}{^k_j}(\breve{e}_i)\breve{e}_k, \quad \breve{\omega}_{kj}=\breve{g}_{kl}\tensor{\breve{\omega}}{^l_j}=-\breve{\omega}_{jk}.
\end{align*}
It gives that \cite{Z1}
\begin{equation}
    \breve{\omega}_{12}=-\frac{\sqrt{1+r^2}}{r}\breve{e}^2, \quad \breve{\omega}_{13}=-\frac{\sqrt{1+r^2}}{r}\breve{e}^3,
    \quad \breve{\omega}_{23}=-\frac{\cot\theta}{r}\breve{e}^3. \label{omega-ij}
\end{equation}

Let $M$ be a spacetime hypersurface in an asymptotically flat Bondi-Sachs spacetime ($\mathscr{L}$, ${\bf g}$), which is given by the inclusion:
\begin{align*}
i\colon M &\longrightarrow \mathscr{L}\\
(y^1,y^2,y^3) &\longmapsto (x^0,x^1,x^2,x^3),
\end{align*}
and, on $\mathscr{L} \backslash \mathscr{L}_c$,
\begin{equation*}
    x^0=u(y^1,y^2,y^3),\quad x^1=y^1=r,\quad x^2=y^2=\theta,\quad x^3=y^3=\phi.
\end{equation*}
Let $g=i^* {\bf g}$ be the induced metric of $M$. For any tangent vectors ${Y}_i$, ${Y}_j \in TM$, $i_* Y_i$, $i_* Y_j$ are the tangent vectors along $M$ in $\mathscr{L}$. Let $e_n$ be the downward unit normal of $M$. The second fundamental form is defined as
\begin{align*}
h(Y_i, Y_j)={\bf g}\left(\nabla_{i_* Y_i} i_* Y_j,e_n\right).
\end{align*}
Now it is straightforward that \cite{HYZ}
\[
i_*\frac{\partial}{\partial y^i}=\frac{\partial x^\mu}{\partial y^i}\frac{\partial}{\partial x^\mu}=\frac{\partial x^0}{\partial y^i}\frac{\partial}{\partial x^0}+\frac{\partial}{\partial x^i}.
\]
Denote $e_i=i_*\breve{e}_i$, $u_{,i}=\frac{\partial u}{\partial y^i}$. Then
\beq
\begin{aligned}\label{eq:eiDef}
    e_1 &=\sqrt{1+r^2}\left(u_{,1}\frac{\partial}{\partial x^0}+\frac{\partial}{\partial x^1}\right), \\
    e_2 &=\frac{1}{r}\left(u_{,2}\frac{\partial}{\partial x^0}+\frac{\partial}{\partial x^2}\right), \\
    e_3 &=\frac{1}{r\sin\theta}\left(u_{,3}\frac{\partial}{\partial x^0}+\frac{\partial}{\partial x^3}\right)
\end{aligned}
\eeq
and
\begin{equation*}
    g(\breve{e}_i,\breve{e}_j)={\bf g}(e_i,e_j), \quad i,\,j=1,\,2,\,3.
\end{equation*}

\begin{definition}\label{def:AsymptoticallyNullDef}
A spacelike hypersurface ($M$, $g$, $h$) in an asymptotically flat Bondi-Sachs spacetime ($\mathscr{L}$, ${\bf g}$) is an asymptotically null initial data set of order $\tau >0$ if, on $\mathscr{L} \backslash \mathscr{L}_c$ with $r$ sufficiently large,
\begin{align*}
g(\breve{e} _i,\breve{e} _j)=\delta_{ij}+a_{ij}, \quad h(\breve{e} _i, \breve{e} _j)=\delta_{ij}+b_{ij},
\end{align*}
where $a _{ij}$, $b _{ij}$ satisfy
\begin{equation*}
    \left\{ a_{ij},\breve{\nabla}_k a_{ij},\breve{\nabla}_l\breve{\nabla}_k a_{ij},
    b_{ij},\breve{\nabla}_kb_{ij} \right\}=O\left(\frac{1}{r^\tau}\right).
\end{equation*}
\end{definition}

Let $(M, g, h)$ be an asymptotically null spacelike hypersurface with the induced metric $g$ and the second fundamental form $h$ in asymptotically flat Bondi-Sachs spacetime ($\mathscr{L}$, ${\bf g}$), which is given by
\begin{equation}\label{eq:uExpand}
    u=u_0+\sqrt{1+r^2}-r+\frac{\left(c^2+d^2+\lambda H^2\right)_{u=u_0}}{12r^3}
    +\frac{a_3(\theta,\phi)}{r^4}
\end{equation}
for sufficiently large $r$ guaranteeing $u>u_0$, where $\lambda$ is a real constant. The induced metric can be obtained by substituting $\mathrm{d}u$ into
\eqref{eq:BondiSachsMetric}. Let $X_n$ be the downward normal vector
\[ X_n=-\frac{\partial}{\partial x^0}-\varrho^i\frac{\partial}{\partial x^i}. \]
Let $e_i$ be given by \eqref{eq:eiDef}. Since $X _n$ is orthogonal to $e
_i$, we obtain
\begin{equation*}
    g(e_i,X_n)=0.
\end{equation*}
We obtain $X_n$ by solving
\begin{equation*}
%\label{eq:varrhoiMatrixEq}
    \begin{pmatrix}
        u_{,1}g_{01}+g_{11} & u_{,1}g_{02}+g_{12} &  u_{,1}g_{03}+g_{13} \\
        u_{,2}g_{01}+g_{21} & u_{,2}g_{02}+g_{22} & u_{,2}g_{03}+g_{23} \\
        u_{,3}g_{01}+g_{31} & u_{,3}g_{02}+g_{32} & u_{,3}g_{03}+g_{33}
    \end{pmatrix}
    \begin{pmatrix}
        \varrho^1 \\
        \varrho^2 \\
        \varrho^3
    \end{pmatrix}
    =-\begin{pmatrix}
        u_{,1}g_{00}+g_{01} \\
        u_{,2}g_{00}+g_{02} \\
        u_{,3}g_{00}+g_{03}
    \end{pmatrix}.
\end{equation*}
Denote the unit normal vector
\begin{equation*}
    e_n=\frac{X_n}{\sqrt{-{\bf g}(X_n,X_n)}}.
\end{equation*}
Then the second fundamental form is given by
\begin{equation*}
    h(\breve{e}_i,\breve{e}_j)={\bf g}\left(\nabla_{e_i}e_j,e_n\right), \quad i,\,j=1,\,2,\,3.
\end{equation*}

\begin{proposition}\label{prop:uExpand}
Suppose ($\mathscr{L}$, ${\bf g}$) is an asymptotically flat Bondi-Sachs spacetime which satisfies the Einstein scalar field equations. Suppose \eqref{eq:gammaExpand}, \eqref{eq:deltaExpand} hold and scalar field $\Psi$ takes expansion \eqref{eq:psiExpand}. Then asymptotically null spacelike hypersurface ($M$, $g$, $h$) given by \eqref{eq:uExpand} for sufficiently large $r$ is an asymptotically null initial data set of order 1
\end{proposition}
\pf On $\mathcal{L} \backslash \mathcal{L}_c$ with $r$ sufficiently large, we expand the following functions at $u=u_0$ by Taylor series
\begin{align*}
    c(u,\theta,\phi) &=c(u_0,\theta,\phi)+c_u(u_0,\theta,\phi)(u-u_0)  \\
    &\quad +\frac{c_{uu}(u_0,\theta,\phi)}{2}(u-u_0)^2+O\left((u-u_0)^3\right), \\
    d(u,\theta,\phi) &=d(u_0,\theta,\phi)+d_u(u_0,\theta,\phi)(u-u_0) \\
    &\quad +\frac{d_{uu}(u_0,\theta,\phi)}{2}(u-u_0)^2+O\left((u-u_0)^3\right), \\
    C(u,\theta,\phi) &=C(u_0,\theta,\phi)+C_u(u_0,\theta,\phi)(u-u_0) \\
    &\quad +\frac{C_{uu}(u_0,\theta,\phi)}{2}(u-u_0)^2+O\left((u-u_0)^3\right), \\
    D(u,\theta,\phi) &=D(u_0,\theta,\phi)+D_u(u_0,\theta,\phi)(u-u_0) \\
    &\quad +\frac{D_{uu}(u_0,\theta,\phi)}{2}(u-u_0)^2+O\left((u-u_0)^3\right), \\
    \mathcal{M}(u,\theta,\phi) &=\mathcal{M}(u_0,\theta,\phi)+\mathcal{M}_u(u_0,\theta,\phi)(u-u_0) \\
    &\quad +\frac{\mathcal{M}_{uu}(u_0,\theta,\phi)}{2}(u-u_0)^2+O\left((u-u_0)^3\right), \\
    N(u,\theta,\phi) &=N(u_0,\theta,\phi)+N_u(u_0,\theta,\phi)(u-u_0) \\
    &\quad +\frac{N_{uu}(u_0,\theta,\phi)}{2}(u-u_0)^2+O\left((u-u_0)^3\right), \\
    P(u,\theta,\phi) &=P(u_0,\theta,\phi)+P_u(u_0,\theta,\phi)(u-u_0) \\
    &\quad +\frac{P_{uu}(u_0,\theta,\phi)}{2}(u-u_0)^2+O\left((u-u_0)^3\right), \\
    H(u,\theta,\phi) &=H(u_0,\theta,\phi)+H_u(u_0,\theta,\phi)(u-u_0) \\
    &\quad +\frac{H_{uu}(u_0,\theta,\phi)}{2}(u-u_0)^2+O\left((u-u_0)^3\right), \\
    K(u,\theta,\phi) &=K(u_0,\theta,\phi)+K_u(u_0,\theta,\phi)(u-u_0) \\
    &\quad +\frac{K_{uu}(u_0,\theta,\phi)}{2}(u-u_0)^2+O\left((u-u_0)^3\right), \\
    L(u,\theta,\phi) &=L(u_0,\theta,\phi)+L_u(u_0,\theta,\phi)(u-u_0) \\
    &\quad +\frac{L_{uu}(u_0,\theta,\phi)}{2}(u-u_0)^2+O\left((u-u_0)^3\right).
\end{align*}
We obtain
\begin{align*}
g(\breve{e}_1,\breve{e}_1)
    =&1-\frac{(1-2\lambda)H^2}{4r^2}+\frac{1}{12r^3} \bigg( 96a_3+6\mathcal{M}  \\
     &+3(l_\theta+l\cot\theta+\hat{l}_\phi\csc\theta) -6(cc_u+dd_u)     \\
     &- 3HH_u -8HK\bigg) +O\left(\frac{1}{r^4}\right),\\
g(\breve{e}_1,\breve{e}_2)
    =&-\frac{l}{2r^2}+\frac{1}{12r^3}\bigg(12N -3l_u -6(cc_\theta+dd_\theta)\\
     &-4c d_\phi\csc\theta +4dc_\phi\csc\theta-8(c^2+d^2)\cot\theta\\
     &-2\lambda HH_\theta  \bigg)+O\left(\frac{1}{r^4}\right),\\
g(\breve{e}_1,\breve{e}_3)
    =&-\frac{\hat{l}}{2r^2}+\frac{1}{12r^3}\bigg( 12P-3\hat{l}_u-6(cc_\phi+dd_\phi)\csc\theta \\
    &+4cd_\theta-4dc_\theta-2\lambda HH_\phi\csc\theta \bigg)+O\left(\frac{1}{r^4}\right),\\
g(\breve{e}_2,\breve{e}_2)
    =&1+\frac{2c}{r}+\frac{2(c^2+d^2)+c_u}{r^2} +\frac{1}{r^3} \bigg(c^3+cd^2\\
     &+2C+2(cc_u +dd_u)+\frac{c_{uu}}{4}\bigg)+O\left(\frac{1}{r^4}\right),\\
g(\breve{e}_2,\breve{e}_3)
    =&\frac{2d}{r}+\frac{d_u}{r^2}
    +\frac{1}{r^3}\bigg(c^2 d+d^3+2D+\frac{d_{uu}}{4}\bigg)
    +O\left(\frac{1}{r^4}\right),    \\
g(\breve{e}_3,\breve{e}_3)
    =&1-\frac{2c}{r}+\frac{2(c^2+d^2)-c_u}{r^2}+\frac{1}{r^3}\bigg(-c^3-cd^2\\
    &-2C+2(cc_u+dd_u)-\frac{c_{uu}}{4}\bigg)+O\left(\frac{1}{r^4}\right),\\
h(\breve{e}_1,\breve{e}_1)
    =&1+\frac{8c^2+8d^2+\left(1+6\lambda \right)H^2}{8r^2}
    +\frac{1}{6r^3}\bigg( 96a_3-6\mathcal{M} \\
    &-3(l_\theta+l\cot\theta+\hat{l}_\phi\csc\theta)+4HK \bigg) +O\left(\frac{1}{r^4}\right),\\
h(\breve{e}_1,\breve{e}_2)
    =&\frac{l}{2r^2}+\frac{1}{12r^3}\bigg( 3l_u+4(c^2+d^2)\cot\theta-24N \\
     &+2(cd_\phi-dc_\phi)\csc\theta-6(cc_\theta+dd_\theta)-8\lambda HH_\theta\bigg)\\
     &+O\left(\frac{1}{r^4}\right),\\
h(\breve{e}_1,\breve{e}_3)
    =&\frac{\hat{l}}{2r^2}+\frac{1}{12r^3}\bigg( 3\hat{l}_u+2(dc_\theta-cd_\theta)-24P \\
     &-6(cc_\phi+dd_\phi)\csc\theta -8\lambda HH_\phi\csc\theta\bigg)+O\left(\frac{1}{r^4}\right),\\
h(\breve{e}_2,\breve{e}_2)
    =&1+\frac{c}{r}+\frac{8c_u+\left(1-2\lambda\right)H^2}{8r^2}
      +\frac{1}{8r^3} \bigg( 6\mathcal{M} -32a_3 \\
     &-8C-4c(c^2+d^2) +10 (cc_u +dd_u)+3c_{uu}-l_\theta\\
     &+3l\cot\theta +3\hat{l}_\phi\csc\theta +(1-2\lambda)cH^2+HH_u\\
     &+\frac{8}{3}HK\bigg)+O\left(\frac{1}{r^4}\right),\\
h(\breve{e}_2,\breve{e}_3)
    =&\frac{d}{r}+\frac{d_u}{r^2}
    +\frac{1}{8r^3}\bigg(  (4\cot^2\theta+4\csc^2\theta+(1-2\lambda)H^2)d\\
    &-4d(c^2+d^2) -8c_\phi\cot\theta\csc\theta-2d_{\phi\phi}\csc^2\theta \\
    &-2d_\theta\cot\theta-2d_{\theta\theta}-8D+3d_{uu}\bigg)+O\left(\frac{1}{r^4}\right),\\
h(\breve{e}_3,\breve{e}_3)
    =&1-\frac{c}{r}+\frac{\left(1-2\lambda\right)H^2-8c_u}{8r^2}+\frac{1}{8r^3}\bigg(6\mathcal{M}-32a_3\\
     &+8C +4c(c^2+d^2)+10(cc_u+dd_u)-3c_{uu}  +3l_\theta \\
     &-l\cot\theta -\hat{l}_\phi\csc\theta-(1-2\lambda)c H^2 +HH_u\\
     &+\frac{8}{3}HK \bigg)+O\left(\frac{1}{r^4}\right).
\end{align*}
Here, all functions in the right hand sides take value at $u=u_0$. Therefore, ($M$, $g$, $h$) is asymptotically null initial data set of order $1$. \qed

%%%%%%%%%%%%%%%%%%%%%%%%%%%%%%%%%%%%%%%%%%%%%%%%%%%%%%%%%%%%%%%%%%%%

\mysection{The Bondi Energy-Momentum and Positivity}
\ls

In this section, we prove the positivity of the Bondi energy-momentum defined by \eqref{eq:newMBondiEnergyMomentumDef} for the Einstein scalar fields when spacetimes are asymptotically flat Bondi-Sachs and the scalar field $\Psi$ satisfies \eqref{eq:psiExpand}. 

By \eqref{eq:R00eqExpand}, we obtain
\begin{equation}\label{eq:BondiLossEnergyMomentum}
    \frac{\mathrm{d}m_\nu}{\mathrm{d}u}=
    -\frac{1}{8\pi}\int_{S^2}\bigg(2c_u^2+2d_u^2+H_u^2\bigg)n^\nu \mathrm{d}S.
\end{equation}
In particular, taking $\nu=0$, \eqref{eq:BondiLossEnergyMomentum} gives rise to the famous Bondi energy loss formula
\begin{equation}\label{eq:BondiLossMass}
    \frac{\mathrm{d}m_0}{\mathrm{d}u}=
    -\frac{1}{8\pi}\int_{S^2}\bigg(2c_u^2+2d_u^2+H_u^2\bigg)\mathrm{d}S \leqslant 0
\end{equation}
for the Einstein scalar field equations. Similar to \cite{HYZ}, we prove the following Bondi energy-momentum loss formula.

\begin{proposition}\label{prop:GeneralizedBondiMassLoss}
Suppose ($\mathscr{L}$, ${\bf g}$) is an asymptotically flat Bondi-Sachs spacetime which satisfies the Einstein scalar field equations. Suppose \eqref{eq:gammaExpand}, \eqref{eq:deltaExpand} hold and scalar field $\Psi$ takes expansion \eqref{eq:psiExpand}. Denote
\begin{align*}
|m|=\sqrt{m_1^2+m_2^2+m_3^2}.
\end{align*}
Then the Bondi energy-momentum for the Einstein scalar fields satisfies
\begin{equation}\label{eq:GeneralizedBondiMassLoss}
    \frac{\mathrm{d}}{\mathrm{d}u}\bigg(m_0-|m|\bigg) \leqslant 0,
\end{equation}
and the equality at some retarded time $u_0$ implies that
\begin{align}
c_u\big|_{u=u_0}=d_u\big|_{u=u_0}=H_u\big|_{u=u_0}=0.  \label{eq:cuduHuZeroatPoint}
\end{align}
\end{proposition}
\pf We use the same argument as Proposition 2.1 \cite{HYZ} and provide the proof here for the sake of completeness.
We assume $|m|\ne 0$, otherwise the proof follows from \eqref{eq:BondiLossMass}.
Using $$(n^1)^2+(n^2)^2+(n^3)^2=1$$ and Cauchy-Schwarz inequality, we obtain
\begin{align*}
\sum\limits_{1\leqslant i\leqslant 3}
&\left( \int_{S^2}\left( 2c_u^2+2d_u^2+H_u^2\right)  n^i \mathrm{d}S \right)^2\\
&=\sum\limits_{1\leqslant i\leqslant 3} \left(\int_{S^2}\big(2c_u^2+2d_u^2+H_u^2 \big)^\frac{1}{2}
\big(2c_u^2+2d_u^2+H_u^2 \big)^\frac{1}{2}  n^i \mathrm{d}S \right)^2\\
&\leqslant \sum\limits_{1\leqslant i\leqslant 3} \int_{S^2}\big(2c_u^2+2d_u^2+H_u^2 \big)\mathrm{d}S
\int_{S^2}\big(2c_u^2+2d_u^2+H_u^2 \big)\big( n^i \big)^2 \mathrm{d}S\\
&=\left(\int_{S^2}\left( 2c_u^2+2d_u^2+H_u^2\right)\mathrm{d}S \right)^2.
\end{align*}
Therefore
\begin{align*}
\sum\limits_{1\leqslant i\leqslant 3} & m_i \int_{S^2}\left( 2c_u^2+2d_u^2+H_u^2\right) n^i \mathrm{d}S \\
    &\leqslant |m|\left[\sum\limits_{1\leqslant i\leqslant 3}\left(
        \int_{S^2}\left( 2c_u^2+2d_u^2+H_u^2\right) n^i \mathrm{d}S \right)^2 \right]^\frac{1}{2}\\
    &\leqslant |m|\int_{S^2}\left( 2c_u^2+2d_u^2+H_u^2\right)\mathrm{d}S.
\end{align*}
This together \eqref{eq:BondiLossEnergyMomentum} yield
\begin{equation*}
\frac{\mathrm{d}}{\mathrm{d}u}\bigg(m_0-|m|\bigg)
=\frac{\mathrm{d}m_0}{\mathrm{d}u}-\frac{1}{|m|}\sum\limits_{i=1}^3 m_i\frac{\mathrm{d}m_i}{\mathrm{d}u}\leqslant 0.
\end{equation*}

If equality holds at some $u_0$ in \eqref{eq:GeneralizedBondiMassLoss}, then there exists constants $\epsilon_i$, which are independent of $\theta$ and $\phi$, such that, for $i=1,\,2,\,3$,
\begin{equation*}
    \epsilon_i \bigg(2c_u^2+2d_u^2+H_u^2\bigg)^\frac{1}{2} =\bigg(2c_u^2+2d_u^2+H_u^2\bigg)^\frac{1}{2} n^i
\end{equation*}
holds at $u_0$. This implies that
\begin{align*}
\bigg(2c_u^2+2d_u^2+H_u^2\bigg)\bigg|_{u=u_0}= 0.
\end{align*}
Therefore \eqref{eq:cuduHuZeroatPoint} holds. \qed

In \cite{Z1} (see also \cite{Z4,Z5}), the second author proved the following positive energy theorem for asymptotically null infinity. 

\begin{theorem} \label{ZX} \cite{Z1, Z4, Z5}
Let ($M$, $g$, $h$) be an asymptotically null initial data set of order $\tau>\frac{3}{2}$ in an asymptotically flat spacetime which satisfies the dominant energy condition
\begin{equation}\label{eq:T00geqSqrt}
    T_{00}\geqslant \sqrt{T_{01}^2+T_{02}^2+T_{03}^2}, \quad
    T_{00}\geqslant |T_{\mu\nu}|
\end{equation}
for any frame $\{e_\alpha \}$ such that $e_0$ is timelike, $e_i$ is spacelike, $1\leqslant i \leqslant 3$. The total energy-momentum of ($M$, $g$, $h$) is given by
\begin{equation*}
%\label{eq:TotalEnergyMomentumDef}
    E_\nu(M)=\frac{1}{16\pi}\lim\limits_{r\to \infty}\int_{S_r}\mathcal{E} n^\nu r\breve{e}^2\wedge \breve{e}^3,
\end{equation*}
where
\begin{equation*}
%\label{eq:mathcalEDef}
    \mathcal{E}=\breve{\nabla}^j a_{1j}
    -\breve{\nabla}_1\operatorname{tr}_{\breve{g}}(a)
    +(a_{22}+a_{33})+2(b_{22}+b_{33}).
\end{equation*}
It holds that
\begin{equation}\label{eq:ADMneq}
    E_0(M)\geqslant \sqrt{\sum\limits_{1\leqslant i\leqslant 3} E_i(M)^2}.
\end{equation}
If $E_0(M)=0$, then the spacetime is flat along $M$.
\end{theorem}

For the Einstein scalar fields,
\begin{align*}
    T_{00} &=\frac{1}{2}\Bigg((e_0(\Psi))^2+\sum\limits_{1\leqslant i\leqslant 3}(e_i(\Psi))^2\Bigg),\\
    T_{ii} &=\frac{1}{2}\Bigg((e_0(\Psi))^2+(e_i(\Psi))^2-\sum\limits_{\substack{1\leqslant j\leqslant 3, j\ne i}}(e_j(\Psi))^2\Bigg),\\
    T_{\alpha i} &=e_\alpha(\Psi) e_i(\Psi), \quad \alpha \neq i.
\end{align*}
Therefore the dominant energy condition \eqref{eq:T00geqSqrt} holds.

\begin{lemma}\label{lemma:mathcalELemma}
Suppose ($\mathscr{L}$, ${\bf g}$) is an asymptotically flat Bondi-Sachs spacetime which satisfies the Einstein scalar field equations. Suppose \eqref{eq:gammaExpand}, \eqref{eq:deltaExpand} hold and scalar field $\Psi$ takes expansion \eqref{eq:psiExpand}. If 
\begin{align*}
c(u_0, \theta, \phi)=d(u_0, \theta, \phi)=0,
\end{align*}
at some retarded time $u=u_0$, then asymptotically null spacelike hypersurface ($M$, $g$, $h$) given by \eqref{eq:uExpand} for sufficiently large $r$ is an asymptotically null initial data set of order 2, and
\begin{equation}\label{eq:mathcalEExpand}
    \mathcal{E}=\frac{4\mathcal{M}(u_0,\theta,\phi)}{r^3}+O\left(\frac{1}{r^4}\right).
\end{equation}
\end{lemma}
\pf Using the asymptotic expansions of $g$, $h$ derived in Proposition \ref{prop:uExpand}, we can see easily that ($M$, $g$, $h$) is of order 2.
Moreover, at $u=u_0$,
\begin{align*}
a_{22}+a_{33} =&\frac{4(c^2+d^2)}{r^2}+\frac{4(cc_u+dd_u)}{r^3}+O\left(\frac{1}{r^4}\right),\\
b_{22}+b_{33} =&\frac{(1-2\lambda)H^2}{4r^2}+\frac{3\mathcal{M}-16a_3+5(cc_u+dd_u)}{2r^3} \\
               &+\frac{l_\theta+l\cot\theta +\hat{l}_\phi\csc\theta}{4r^3}+\frac{H (3H_u+8K)}{12r^3}
                +O\left(\frac{1}{r^4}\right).
\end{align*}
Using the formula
\begin{align*}
    \breve{\nabla}_k a_{ij}
   =\breve{e}_k(a_{ij})-\sum\limits_{l=1}^3a_{jl} \breve{\omega}_{li}(\breve{e}_k)
    -\sum\limits_{l=1}^3 a_{il}\breve{\omega}_{lj}(\breve{e}_k),
\end{align*}
we obtain
\begin{align*}
\mathcal{E} =&\breve{\nabla}^j a_{1j}-\breve{e}_1\left(\operatorname{tr}_{\breve{g}}(a)\right)+a_{22}+a_{33}+2(b_{22}+b_{33}) \\
            =&\breve{e}_1(a_{11})+\breve{e}_2(a_{12})+\breve{e}_3(a_{13})
              -\sum\limits_{j=1}^3 \left( a_{j2}\breve{\omega}_{21}(\breve{e}_j)+a_{j3}\breve{\omega}_{31}(\breve{e}_j)\right)  \\
             &-a_{11}\sum\limits_{j=1}^3 \breve{\omega}_{1j}(\breve{e}_j)
              -a_{12}\sum\limits_{j=1}^3 \breve{\omega}_{2j}(\breve{e}_j)
              -a_{13}\sum\limits_{j=1}^3 \breve{\omega}_{3j}(\breve{e}_j) \\
            &-\breve{e}_1(a_{11})-\breve{e}_1(a_{22})-\breve{e}_1(a_{33})+a_{22}+a_{33} +2(b_{22}+b_{33}) \\
           =&\frac{1}{r}\frac{\partial a_{12}}{\partial \theta}+\frac{1}{r\sin\theta}\frac{\partial a_{13}}{\partial \phi}
             -\sum\limits_{j=1}^3 a_{j2}\frac{\sqrt{1+r^2}}{r}\breve{e}^2(\breve{e}_j)-\sum\limits_{j=1}^3 a_{j3}\frac{\sqrt{1+r^2}}{r}\breve{e}^3(\breve{e}_j)\\
            &+\frac{\sqrt{1+r^2}}{r}a_{11}\breve{e}^2(\breve{e}_2)+\frac{\sqrt{1+r^2}}{r}a_{11}\breve{e}^3(\breve{e}_3)
             -a_{12}\frac{\sqrt{1+r^2}}{r}\breve{e}^2(\breve{e}_1)\\
            &+\frac{\cot\theta}{r}a_{12}\breve{e}^3(\breve{e}_3)-a_{13}\frac{\sqrt{1+r^2}}{r}\breve{e}^3(\breve{e}_1)
             -\frac{\cot\theta}{r}a_{13}\breve{e}^3(\breve{e}_2)\\
            &-\sqrt{1+r^2}\frac{\partial a_{22}}{\partial r}-\sqrt{1+r^2}\frac{\partial a_{33}}{\partial r}+a_{22}+a_{33}+2(b_{22}+b_{33}) \\
           =&\frac{1}{r}\frac{\partial a_{12}}{\partial \theta}+\frac{1}{r\sin\theta}\frac{\partial a_{13}}{\partial \phi}
             -\frac{\sqrt{1+r^2}}{r}(a_{22}+a_{33})+\frac{2\sqrt{1+r^2}}{r}a_{11}\\
            &+\frac{\cot\theta}{r}a_{12}-\sqrt{1+r^2}\frac{\partial (a_{22}+a_{33})}{\partial r}
             +a_{22}+a_{33}+2(b_{22}+b_{33})\\
          =&\frac{8\sqrt{1+r^2}}{r^3}\left(c^2+d^2\right)+\frac{12\sqrt{1+r^2}}{r^4}\left(cc_u+dd_u\right)\\
           &+\frac{\mathcal{M}+16a_3-cc_u-dd_u}{r^3}+\frac{3\mathcal{M}-16a_3+5(cc_u+dd_u)}{r^3}\\
           &+\frac{l_\theta+l\cot\theta+\hat{l}_\phi\csc\theta}{2r^3}+O\left(\frac{1}{r^4}\right)  \\
          =&\frac{8(c^2+d^2)}{r^2}+\frac{8\mathcal{M}+32(cc_u+dd_u)+l_\theta+l\cot\theta +\hat{l}_\phi\csc\theta}{2r^3}
            +O\left(\frac{1}{r^4}\right).
\end{align*}
If $c(u_0, \theta, \phi)=d(u_0, \theta, \phi)=0$, we have, at $u=u_0$,
\begin{align*}
l_\theta+l\cot\theta+\hat{l}_\phi\csc\theta
    =&-2c(u_0,\theta,\phi)+c_{\theta\theta}(u_0,\theta,\phi)-c_{\phi\phi}(u_0,\theta,\phi)\csc^2\theta \\
     &+2d_{\theta\phi}(u_0,\theta,\phi)\csc\theta +3c_\theta(u_0,\theta,\phi)\cot\theta\\
     &+2d_\phi(u_0,\theta,\phi)\cot\theta\csc\theta =0.
\end{align*}
Therefore, \eqref{eq:mathcalEExpand} follows. \qed

\begin{theorem}\label{main thm}
Suppose ($\mathscr{L}$, ${\bf g}$) is an asymptotically flat Bondi-Sachs spacetime which satisfies the Einstein scalar field equations. Suppose \eqref{eq:gammaExpand}, \eqref{eq:deltaExpand} hold and scalar field $\Psi$ takes expansion \eqref{eq:psiExpand}. If 
\begin{align*}
c(u_0, \theta, \phi)=d(u_0, \theta, \phi)=0,
\end{align*}
at some retarded time $u=u_0$, then,

(i) For all $u\leqslant u_0$,
\begin{equation}\label{eq:positiveBondimass}
    m_0(u)\geqslant \sqrt{\sum\limits_{1\leqslant i\leqslant 3}m_i^2(u)}.
\end{equation}
If $m_0(u_0)=0$, the spacetime is flat and the scalar field $\Psi$ vanishes along any asymptotically null spacelike hypersurface given by
\eqref{eq:uExpand} for sufficiently large $r$.

(ii) If there exists $u_1<u_0$ such that the equality holds at $u=u_0, \,u_1$ in \eqref{eq:positiveBondimass}, then the equality holds and
\begin{align}
c=d=H_u =0 \label{cdHu}
\end{align}
for all $u \in [u_1,u_0]$.

(iii) If there exists $u_1<u_0$ such that $m_0(u_1)=m_0(u_0)=0$, then the spacetime is flat and the scalar field $\Psi$ vanishes for all $u\in (u_1,u_0]$.
\end{theorem}
\pf (i) Choose an asymptotically null spacelike hypersurface $\breve{M}$ given by \eqref{eq:uExpand} for sufficiently large $r$. By Lemma \ref{lemma:mathcalELemma}, we obtain, for $\nu=0, \,1, \,2, \,3$,
\begin{align*}
E_{\nu}(M)&=\frac{1}{16\pi}\lim\limits_{r\to \infty}\int_{S_r}\mathcal{E}n^\nu r\breve{e}^2\wedge \breve{e}^3 \\
                  &=\frac{1}{16\pi}\lim\limits_{r\to \infty}\int_{S^2}\mathcal{E} n^\nu r r^2\sin\theta\mathrm{d}\theta\mathrm{d}\phi \\
                  &=\frac{1}{16\pi}\lim\limits_{r\to \infty}\int_{S^2}\mathcal{E}n^\nu r^3\mathrm{d}S \\
                  &=\frac{1}{4\pi}\lim\limits_{r\to \infty}\int_{S^2}\frac{\mathcal{M}(u_0,\theta,\phi)}{r^3} n^\nu r^3\mathrm{d}S \\
                  &=m_\nu(u_0).
\end{align*}
As the dominant energy condition holds, we can apply \eqref{eq:ADMneq} and obtain
\begin{equation*}
m_0(u_0)=E_0(M)\geqslant \sqrt{\sum\limits_{1\leqslant i\leqslant 3} E ^2 _i(M)}=\sqrt{\sum\limits_{1\leqslant i\leqslant 3}m_i^2(u_0)}.
\end{equation*}
Then the Bondi energy-momentum loss formula \eqref{eq:GeneralizedBondiMassLoss} gives \eqref{eq:positiveBondimass}.
\begin{align*}
m_0(u)-\sqrt{\sum\limits_{1\leqslant i\leqslant 3}m_i^2(u)}
    \geqslant
    m_0(u_0)-\sqrt{\sum\limits_{1\leqslant i\leqslant 3}m_i^2(u_0)}\geqslant 0.
\end{align*}
This gives \eqref{eq:positiveBondimass} for all $u\leqslant u_0$. If $m_0(u_0)=0$, then $E_0(M)=0$ and $T_{00}=0$ along any asymptotically null spacelike hypersurface given by \eqref{eq:uExpand} for sufficiently large $r$. This implies that the spacetime is flat and the scalar field $\Psi$ vanishes along these hypersurfaces.

(ii) By the Bondi energy-momentum loss formula \eqref{eq:GeneralizedBondiMassLoss}, we know that the equality and \eqref{eq:cuduHuZeroatPoint} hold for all $u \in [u_1,\, u_0]$. Then \eqref{cdHu} holds as $c=d=0$ at $u=u_0$.

(iii) By the Bondi energy loss formula \eqref{eq:BondiLossMass}, we know that $m_0(u)=0$ and \eqref{cdHu} holds for all $u \in [u_1,\, u_0]$. By (i), the spacetime is flat and the scalar field $\Psi$ vanishes along any asymptotically null spacelike hypersurface given by
\begin{equation*}
    u=\check{u}_0+\sqrt{1+r^2}-r+\frac{\left(c^2+d^2+\lambda H^2\right)_{u=\check{u}_0}}{12r^3}
    +\frac{a_3(\theta,\phi)}{r^4}
\end{equation*}
for sufficiently large $r$ guaranteeing $u>\check{u}_0$, where $u_1 \leqslant \check{u}_0 \leqslant u_0$. This gives the conclusion. \qed

\section*{Appendix A: Explicit Formulas for $\mathscr{R}_1,\ldots,\mathscr{R}_7$ in \eqref{eq:EQ1}-\eqref{eq:EQ7} }
\begin{align*}
\mathscr{R}_1 =& \frac{1}{2}r\Big(\gamma_r^2\cosh^2(2\delta)+\delta_r^2\Big)+\frac{1}{4}r\Psi_r^2,  \\
\mathscr{R}_2 =& 2r^2\Big(\beta_{r\theta}+2\delta_r\delta_\theta-2r^{-1}\beta_\theta
	            -4\gamma_r\delta_\theta\sinh(2\delta)\cosh(2\delta) \\
	           &-\big(\gamma_{r\theta}-2\gamma_r\gamma_\theta+2\gamma_r\cot\theta\big)\cosh^2(2\delta) \Big)\\
	           &+2r^2\mathrm{e}^{2\gamma}\csc\theta\Big(-\delta_{r\phi}
                   -2\delta_r\gamma_\phi+(\gamma_{r\phi}+2\gamma_r\gamma_\phi)\sinh(2\delta)\cosh(2\delta) \\
               &+2\gamma_r\delta_\phi(1+2\sinh^2(2\delta))\Big)+2r^2\Psi_r\Psi_\theta,\\
\mathscr{R}_3 =& 2r^2\mathrm{e}^{-2\gamma}\Big(-\delta_{r\theta}+2\delta_r\gamma_\theta-2\delta_r\cot\theta
               -(\gamma_{r\theta}-2\gamma_r\gamma_\theta\\
               &+2\gamma_r\cot\theta)\cosh(2\delta)\sinh(2\delta)-2\gamma_r\delta_\theta(1+2\sinh^2(2\delta)) \Big) \\
               &+2r^2\csc\theta\Big(\beta_{r\phi}+2\delta_r\delta_\phi-2r^{-1}\beta_\phi
	            +4\gamma_r\delta_\phi\cosh(2\delta)\sinh(2\delta) \\
               &+(\gamma_{r\phi}+2\gamma_r\gamma_\phi)\cosh^2(2\delta) \Big)+2 r^2 \Psi_r\Psi_\phi\csc\theta,\\
\mathscr{R}_4 =& 2\mathrm{e}^{2\beta}\csc\theta\Big((\beta_{\theta\phi}+\beta_\theta\beta_\phi+2\delta_\theta\delta_\phi)\sinh(2\delta) \\
	           &+(\delta_{\theta\phi}+\delta_\phi\cot\theta+\delta_\theta\gamma_\phi-\gamma_\theta\delta_\phi
                +\delta_\theta\beta_\phi+\beta_\theta\delta_\phi)\cosh(2\delta) \Big) \\
               &-\mathrm{e}^{2\beta-2\gamma}\Big( (\beta_{\theta\theta}+\beta_\theta^2+\beta_\theta\cot\theta+2\gamma_\theta^2+2\delta_\theta^2
                -1-\gamma_{\theta\theta}\\
               &-3\gamma_\theta\cot\theta -2\beta_\theta\gamma_\theta)\cosh(2\delta)
                +(\delta_{\theta\theta}+3\delta_\theta\cot\theta+2\beta_\theta\delta_\theta\\
               &-4\gamma_\theta\delta_\theta)\sinh(2\delta) \Big)
                -\mathrm{e}^{2\beta+2\gamma}\csc^2\theta\Big((\beta_{\phi\phi}+\beta_\phi^2+2\gamma_\phi^2+2\delta_\phi^2\\
               &+\gamma_{\phi\phi}+2\beta_\phi\gamma_\phi)\cosh(2\delta)
                +(\delta_{\phi\phi}+2\beta_\phi\delta_\phi+4\gamma_\phi\delta_\phi)\sinh(2\delta) \Big) \\
	           &-\frac{1}{4}r^4\mathrm{e}^{-2\beta}\Big( (\mathrm{e}^{2\gamma}U_r^2
                +\mathrm{e}^{-2\gamma}W_r^2)\cosh(2\delta)+2U_r W_r \sinh(2\delta) \Big) \\
	           &+\frac{1}{2}r( rU_{r\theta}+rU_r\cot\theta+4U_\theta+4U\cot\theta)
                +\frac{1}{2}r\csc\theta (r W_{r\phi}+4W_\phi) \\
               &-\frac{1}{2}\mathrm{e}^{2\beta}\Big((\mathrm{e}^{-2\gamma}\Psi_\theta^2+\mathrm{e}^{2\gamma}\Psi_\phi^2\csc^2\theta)\cosh(2\delta)
                -2\Psi_\theta\Psi_\phi \sinh(2\delta)\csc\theta\Big),\\
\mathscr{R}_5 = &\frac{1}{2}(\gamma_r V_r+\gamma_{rr}V+r^{-1}\gamma_r V)
	             \cosh(2\delta)+2\gamma_r\delta_r V\sinh(2\delta) \\
	            &+\frac{1}{8}r^3\mathrm{e}^{-2\beta}(\mathrm{e}^{2\gamma}U_r^2-\mathrm{e}^{-2\gamma}W_r^2)
	             +\frac{1}{2}r^{-1}\mathrm{e}^{2\beta-2\gamma}(\beta_{\theta\theta}+\beta_\theta^2-\beta_\theta\cot\theta) \\
	            &-\frac{1}{2}r^{-1}\mathrm{e}^{2\beta+2\gamma}(\beta_{\phi\phi}+\beta_\phi^2)\csc^2\theta
	             +r^{-1}\mathrm{e}^{2\beta}(\beta_\theta\delta_\phi-\beta_\phi\delta_\theta)\csc\theta \\
	            &+\frac{1}{4}r\mathrm{e}^{2\gamma}\csc\theta
                  \Big( (U_{r\phi}+2r^{-1}U_\phi)\sinh(2\delta)+4\delta_r U_\phi\cosh(2\delta) \Big) \\
                &-\frac{1}{4}r\mathrm{e}^{-2\gamma}\Big( (W_{r\theta}-W_r\cot\theta)\sinh(2\delta)
                 +2r^{-1}(W_\theta-W\cot\theta)\sinh(2\delta) \\
	            &+4\delta_r(W_\theta-W\cot\theta)\cosh(2\delta) \Big)
	             -\frac{1}{4}r\big( U_{r\theta}+2r^{-1}U_\theta-U_r\cot\theta\\
                &-2r^{-1}U\cot\theta+4r^{-1}\gamma_\theta U+4\gamma_{r\theta}U+2\gamma_\theta U_r+2\gamma_rU_\theta\\
                &+2\gamma_r U\cot\theta \big)\cosh(2\delta)
	             -r(\delta_r U_\theta+2\gamma_r\delta_\theta U+2\delta_r\gamma_\theta U\\
                &-\delta_r U\cot\theta)\sinh(2\delta)
	             +\frac{1}{4}r\csc\theta( W_{r\phi}+2r^{-1}W_\phi-4r^{-1}\gamma_\phi W\\
                &-4\gamma_{r\phi}W-2\gamma_\phi W_r-2\gamma_r W_\phi)\cosh(2\delta)+r\csc\theta(\delta_r W_\phi
                 -2\delta_r\gamma_\phi W\\
                &-2\gamma_r\delta_\phi W)\sinh(2\delta)
                 +\frac{1}{4r}\mathrm{e}^{2\beta}\left(\mathrm{e}^{-2\gamma}\Psi_\theta^2-\mathrm{e}^{2\gamma}\Psi_\phi^2\csc^2\theta \right),\\
\mathscr{R}_6 = &\frac{1}{2}\Big(\delta_r V_r+\delta_{rr}V+r^{-1}\delta_r V-2\gamma_r^2 V\cosh(2\delta)\sinh(2\delta)\Big)
	             -\frac{1}{2r}\mathrm{e}^{2\beta-2\gamma}\\
                &(\beta_{\theta\theta}+\beta_\theta^2-\beta_\theta\cot\theta)\sinh(2\delta)
                 -\frac{1}{2r}\mathrm{e}^{2\beta+2\gamma}\csc^2\theta(\beta_{\phi\phi}+\beta_\phi^2)\sinh(2\delta) \\
	            &-\frac{1}{r}\mathrm{e}^{2\beta}\csc\theta\big( -\beta_{\theta\phi}-\beta_\theta\beta_\phi+\beta_\phi\cot\theta
                 +\beta_\theta\gamma_\phi-\gamma_\theta\beta_\phi \big)\cosh(2\delta) \\
	            &+\frac{1}{8}r^3\mathrm{e}^{-2\beta}\Big((\mathrm{e}^{2\gamma}U_r^2+\mathrm{e}^{-2\gamma}W_r^2)\sinh(2\delta)
                 +2U_r W_r\cosh(2\delta) \Big) \\
                &-\frac{1}{2}r\Big( 2\delta_{r\theta}U+\frac{2}{r}\delta_\theta U+\delta_r U_\theta+\delta_\theta U_r+\delta_r U\cot\theta
                 -2(\gamma_r U_\theta\\
                &-\gamma_r U\cot\theta+2\gamma_r\gamma_\theta U)\cosh(2\delta)\sinh(2\delta) \Big)
                 -\frac{1}{2}r\csc\theta\Big( 2\delta_{r\phi}W\\
                &+\frac{2}{r}\delta_\phi W +\delta_r W_\phi+\delta_\phi W_r +2(\gamma_r W_\phi-2\gamma_r\gamma_\phi W)\cosh(2\delta)\sinh(2\delta)\Big) \\
                & -\frac{1}{4}r\mathrm{e}^{-2\gamma}\Big( W_{r\theta}-W_r\cot\theta+ \frac{2}{r}(W_\theta-W\cot\theta)-4\gamma_r(W_\theta-W\cot\theta)\\
                &\cosh^2(2\delta)\Big)-\frac{1}{4}r\mathrm{e}^{2\gamma}\csc\theta
	             \Big(U_{r\phi}+\frac{2}{r}U_\phi+4\gamma_rU_\phi\cosh^2(2\delta) \Big) \\
                &-\frac{\mathrm{e}^{2\beta}}{4r}\Big((\mathrm{e}^{-2\gamma}\Psi_\theta^2+\mathrm{e}^{2\gamma}\Psi_\phi^2\csc^2\theta)\sinh(2\delta)
                 -2\Psi_\theta\Psi_\phi\cosh(2\delta)\csc\theta \Big),\\
\mathscr{R}_7 = &\frac{1}{2}V\Psi_{rr}-rU\Psi_{r\theta}-rW\csc\theta\Psi_{r\phi}
                 +\frac{1}{2r}\mathrm{e}^{2\beta-2\gamma}\cosh(2\delta)\Psi_{\theta\theta} \\
                &-\frac{1}{r}\mathrm{e}^{2\beta}\Psi_{\theta\phi}\csc\theta\sinh(2\delta)
                 +\frac{1}{2r}\mathrm{e}^{2\beta+2\gamma}\Psi_{\phi\phi}\cosh(2\delta)\csc^2\theta \\
	            &+\frac{r}{2}\Big(\frac{V}{r^2}+\frac{V_r}{r}-U\cot\theta-U_\theta-W_\phi\csc\theta\Big)\Psi_r
                 +\bigg( \frac{1}{2r}\mathrm{e}^{2\beta-2\gamma}\\
                &\Big(\cosh(2\delta)\cot\theta
                 +2\cosh(2\delta)\beta_\theta+2\sinh(2\delta)\delta_\theta-2\cosh(2\delta)\gamma_\theta\Big)\\
                &-\frac{1}{2}(2U+rU_r)
                 -\frac{1}{r}\mathrm{e}^{2\beta}\cosh(2\delta)\csc\theta\delta_\phi 
                 -\frac{1}{r}\mathrm{e}^{2\beta}\sinh(2\delta)\csc\theta\beta_\phi \bigg)\Psi_\theta\\
                &+\bigg(\frac{1}{2r} \mathrm{e}^{2\beta+2\gamma}\csc^2\theta\Big( 2\cosh(2\delta)\gamma_\phi
                 +2\sinh(2\delta)\delta_\phi+2\cosh(2\delta)\beta_\phi\Big)\\
                &-\frac{1}{2}(rW_r+2W)\csc\theta
                 -\frac{1}{r}\mathrm{e}^{2\beta}\beta_\theta\sinh(2\delta)\csc\theta\\
                &-\frac{1}{r}\mathrm{e}^{2\beta}\delta_\theta\cosh(2\delta)\csc\theta \bigg)\Psi_\phi.
\end{align*}

\section*{Appendix B: Explicit Formulas for Higher-Order coefficients}

The following functions are explicit formulas for certain higher order coefficients in asymptotic expansions in Sect. 3.
\begin{align*}
\overset{4}{\beta} =&\frac{1}{8}\bigg( ( c^2+d^2)^2-6(cC+dD)-3HL-2K^2 \bigg),\\
\overset{4}{U} =&\frac{1}{24}\bigg(-48c^3\cot\theta-12c^2(3c_\theta+2d_\phi\csc\theta)-24 d^2 c_\theta \\
                &-6c( 2dc_\phi\csc\theta+8d^2\cot\theta+2dd_\theta+12N-H^2\cot\theta)\\
                &+36 C_\theta +72C\cot\theta-36d^2 d_\phi\csc\theta -72 d P +36D_\phi\csc\theta\\
                &+3 H^2 c_\theta +3H^2d_\phi\csc\theta+8 HK_\theta-4KH_\theta\bigg),\\
\overset{4}{W} =&\frac{1}{24}\bigg(-12c^2(-3c_\phi\csc\theta+2d_\theta+4d\cot\theta)-12d^2(3d_\theta\\
                &-2c_\phi\csc\theta)-12c(dc_\theta- dd_\phi\csc\theta-6P)+36(-C_\phi\csc\theta \\
                &+D_\theta+2D\cot\theta)-48 d^3\cot\theta -72 dN -3 H^2(c_\phi\csc\theta-d_\theta)\\
                &-4 KH_\phi\csc\theta+8 HK_\phi\csc\theta +6 dH^2\cot\theta \bigg),\\
\overset{1}{V} =&\frac{1}{12}\bigg(2(25\cos^2\theta-1) c^2\csc^2\theta +2c(5c_{\phi\phi}\csc^2\theta+37c_\theta\cot\theta\\
                &+5c_{\theta\theta}+24 d_\phi\cot\theta\csc\theta)+2d(-24 c_\phi \cot\theta\csc\theta +37 d_\theta\cot\theta \\
                &+5d_{\theta\theta}+5d_{\phi\phi}\csc^2\theta )-32c_\phi d_\theta\csc\theta +32 c_\theta d_\phi\csc\theta +22 c_\theta^2\\
                &+22c_\phi^2\csc^2\theta +2(24\csc^2\theta-25)d^2+22d_\phi^2\csc^2\theta+22 d_\theta^2\\
                &-12 N_\theta-12 N\cot\theta-12 P_\phi\csc\theta +3  H^2-3 H( H_{\theta\theta}\\
                &+ H_\theta\cot\theta+ H_{\phi\phi}\csc^2\theta )+3 H_\theta^2+3H_\phi^2 \csc^2\theta\bigg).\\
\overset{3}{\gamma} _u =& \frac{1}{48}\bigg( 2c\big( 11c_{\theta\theta}+21c_\theta \cot\theta-11c_{\phi\phi}\csc^2\theta-48 d_u d +16 d_{\theta\phi}\csc\theta\\
    &+8 d_\phi\cot\theta\csc\theta+12 \mathcal{M}\big)-\big(10c_\phi^2+10d_\phi^2-3 H_\phi^2 +3 H H_{\phi\phi}\\
    &-12P_\phi\sin\theta\big)\csc^2\theta+10\big(c_\theta^2+d_\theta^2\big)+2d\big( -16 c_{\theta\phi}\csc\theta \\
    &-8c_\phi\cot\theta\csc\theta+11d_{\theta\theta}+21 d_\theta \cot\theta-11 d_{\phi\phi}\csc^2\theta \big) \\
    &-8(3-\cos(2\theta))c^2\csc^2\theta-16d^2 (6c_u+1+\csc^2\theta)+3 H H_{\theta\theta}\\
    &-3 H_\theta(H_\theta+H\cot\theta)+12N\cot\theta-12N_\theta \bigg),\\
\overset{3}{\delta} _u =&\frac{1}{24}\bigg( 2c\big(11c_{\theta\phi}\csc\theta
    +5c_\phi\cot\theta\csc\theta+24c_u d-4d_{\theta\theta}-8d_\theta\cot\theta \\
    &+4d_{\phi\phi}\csc^2\theta \big)+\big( 10c_\theta c_\phi+10d_\theta d_\phi 
     +3( H H_{\theta\phi}-H_\theta H_\phi \\
    &-HH_\phi\cot\theta)-6N_\phi \big)\csc\theta+2d\big( 4c_{\theta\theta}+8c_\theta\cot\theta-4c_{\phi\phi}\csc^2\theta\\
    &+11 d_{\theta\phi}\csc\theta+5 d_\phi\cot\theta\csc\theta+6 \mathcal{M}\big)-6P_\theta+6P\cot\theta \bigg).\\
\overset{3}{\Psi} _u
    =&\frac{1}{4}\bigg( 4c_\theta H_\theta-4 c_\phi H_\phi\csc^2\theta+2Hc_{\theta\theta}+6 Hc_\theta\cot\theta -2 Hc_{\phi\phi}\csc^2\theta \\
     &+2c\big(H_{\theta\theta}+3H_\theta\cot\theta -H_{\phi\phi}\csc^2\theta-2H\big)+4 d_\theta H_\phi\csc\theta \\
     &+4d_\phi H_\theta\csc\theta +4 Hd_{\theta\phi}\csc\theta +4H d_\phi\cot\theta\csc\theta +4 dH_{\theta\phi}\csc\theta \\
     &+4 dH_{\phi}\cot\theta\csc\theta+2 H \mathcal{M} - K_{\theta\theta}-K_\theta \cot\theta- K_{\phi\phi}\csc^2\theta -2 K \bigg).
\end{align*}   

\bigskip

\footnotesize {

\noindent {\bf Acknowledgement} The authors are grateful to the referees for many valuable suggestions. The work is supported by the National Natural Science Foundation of China 12326602, the special foundation for Junwu and Guangxi Ba Gui Scholars.

\noindent {\bf Conflict of interest} The authors have no conflict of interest to declare that are relevant to the content of this article.

}

%%%%%%%%%%%%%%%%%%%%%%%%%%%%%%%%%%%%%%%%%%%%%%%%%%%%%%%%%%%%%%%%%%%%%%%

\end{document}